\documentclass[prd,preprint,tightenlines,showpacs]{revtex4}
\RequirePackage{amsfonts}
\RequirePackage{amssymb}
\RequirePackage{amsmath}
\RequirePackage{graphicx}
\RequirePackage{makeidx}
\RequirePackage{ifthen}
\RequirePackage{color}
\newcommand{\be}{\begin{eqnarray}}          \newcommand{\ee}{\end{eqnarray}}
\newcommand{\ba}{\begin{align}}          \newcommand{\ea}{\end{align}}
\newcommand{\benum}{\begin{enumerate}}   
\newcommand{\eenum}{\end{enumerate}}
\newcommand{\bitem}{\begin{itemize}}          
\newcommand{\eitem}{\end{itemize}}

\newcommand{\abs}[1]{\ensuremath{\left| #1 \right|}}   % absolute value %
\definecolor{myblue}{rgb}{.1,.1,.7}
\definecolor{dcyan}{rgb}{.0,.6,.6}
\definecolor{dmagenta}{rgb}{0.6,0.0,0.6}
\definecolor{brown}{rgb}{0.6,0.2,0.}
\definecolor{darkblue}{rgb}{0,0,0.6}
\definecolor{darkred}{rgb}{0.75,0.0,0.0}
\definecolor{darkgreen}{rgb}{0.0,0.6,0.0}

\newcommand{\blue}{\color{blue}}

\newcommand{\usemarker}{Y}
\newcommand{\marker}[1]{
       \ifthenelse{\equal{\usemarker}{Y}}
                     {\mbox{}\marginpar{\tt #1}}{}
               }

\newcommand{\mk}[1]{
\ifthenelse{\equal{\usemarker}{Y}}
{\noindent\hskip -1truecm {\bf\blue$^{#1}$}}{}
}

\newcommand{\tmop}[1]{#1}
\newcommand{\tmfloatcontents}{}
\newlength{\tmfloatwidth}
\newcommand{\tmfloat}[5]{
  \renewcommand{\tmfloatcontents}{#4}
  \setlength{\tmfloatwidth}{\widthof{\tmfloatcontents}+1in}
  \ifthenelse{\equal{#2}{small}}
    {\ifthenelse{\lengthtest{\tmfloatwidth > \linewidth}}
      {\setlength{\tmfloatwidth}{\linewidth}}{}}
    {\setlength{\tmfloatwidth}{\linewidth}}  \begin{minipage}[#1]{\tmfloatwidth}
    \begin{center}
      \tmfloatcontents
      \captionof{#3}{#5}
    \end{center}
  \end{minipage}}
\newenvironment{itemizedot}
  {\begin{itemize}}{\end{itemize}}
\begin{document}
\title{Charmless decays $B\to \pi\pi$, $\pi K$ and $K K$ in broken $SU(3)$ symmetry\\
%with subleading flavor topological diagrams
}
\author{Yue-Liang Wu}
\affiliation{Institute of Theoretical Physics, CAS, Beijing, 100080, China}
\email[Email: ]{ylwu@itp.ac.cn}
\author{Yu-Feng Zhou}
\affiliation{Institute for Physics, Dortmund University,D-44221,Dortmund, Germany}
\email[Email: ]{zhou@zylon.physik.uni-dortmund.de}

%\date{$Date: 2005/03/07 14:17:56 $}
\begin{abstract}
  Charmless $B$ decay modes $B \rightarrow \pi \pi, \pi K$ and $KK$ are
  systematically investigated with and without flavor SU(3) symmetry.
  Independent analyses on $\pi \pi$ and $\pi K$ modes $both$ favor a large ratio
  between color-suppressed tree ($C$) and tree ($T )$ diagram, which suggests
  that they are more likely to originate from long distance effects.
  The sizes of QCD penguin diagrams extracted individually from  $\pi\pi$,
  $\pi K$ and $KK$ modes are found to follow a pattern of SU(3) breaking in
  agreement with the naive factorization estimates.
  Global fits to these modes are done under various scenarios of SU(3)
  relations. The results show good determinations of weak phase $\gamma$ in
  consistency with the Standard Model (SM), but a large electro-weak penguin $(
  P_{\tmop{EW}} )$ relative to $T + C$ with a large relative strong phase are
  favored, which requires an big enhancement of color suppressed electro-weak
  penguin ($P_{\tmop{EW}}^C$) compatible in size but destructively interfering
  with $P_{\tmop{EW}}$ within the SM, or implies new physics.
  Possibility of sizable contributions from nonfactorizable diagrams
  such as $W$-exchange ($E$), annihilation ($A$) and
  penguin-annihilation diagrams($P_A$) are investigated.
  The implications to the branching ratios and CP violations in $K K$
  modes are discussed.
\end{abstract}
%\preprint{hep-ph/xxx}
\preprint{DO-TH 05/02}

%\keywords{}

%%%%%%%%%%%%%%%%%%%%%%%%%%%%%%%%%%%%%%%%%%%%%%%%%%%%
\maketitle
%%%%%%%%%%%%%%%%%%%%%%%%%%%%%%%%%%%%%%%%%%%%%%%%%%%%%%%%%%%%%%%%%%%%%%%%%%%%%%%%
%%%%%%%%%%%%%%%%%%%%%%%%%%%%-Begin-%%%%%%%%%%%%%%%%%%%%%%%%%%%%%%%%%%%%%%%%%%%%%
%%%%%%%%%%%%%%%%%%%%%%%%%%%%%%%%%%%%%%%%%%%%%%%%%%%%%%%%%%%%%%%%%%%%%%%%%%%%%%%%

%S1
\section{introduction}

In the recent years, the two $B$-factories have succeed in
steadily improving the measurements of hadronic charmless $B$
decays. At present, all the branching ratios of $B \rightarrow \pi
\pi$ and $\pi K$ modes have been measured with good accuracy. The
large direct CP violations have also been established in $\pi^+
\pi^-$ and $\pi^+ K^-$ modes \cite{ICHEP04data}. Their
implications have been reported in a recent short
paper\cite{Wu:2004xx}. It has been shown that the weak phase
$\gamma$ can well be determined to be consistent with the standard
model, it prefers a relative large electroweak penguin with a
large strong phase and also favors an enhanced color-suppressed
tree diagram. In this longer paper, we shall provide a more
detailed analysis with including subleading diagrams and their
implications to $K K$ modes as well as paying attention to SU(3)
broken effects.

 It is of interest to note that the signs of the direct CP
violations, if confirmed by the future experiments would agree
with the results from perturbative QCD approach
\cite{Keum:2000ph,Keum:2002vi} while posse a challenge to the
naive factorization\cite{Ali:1998nh,Ali:1998eb} and QCD
factorization calculations
\cite{Beneke:1999br,Beneke:2000ry,Beneke:2001ev,Beneke:2003zv}.
These impressive new results have triggered a large amount of
theoretical efforts to understand the strong interaction dynamics
of those decays
\cite{Barshay:2004hb,He:2004ck,Charng:2004ed,Cheng:2004ru,Carruthers:2004gj},
extract the weak phases in the Cabbibo-Kobayashi-Maskawa (CKM)
matrix \cite{Ali:2004hb,Parkhomenko:2004mn} and explore new
physics
\cite{Buras:2003dj,Buras:2004ub,Buras:2004th,Barger:2004hn,He:2004wh,London:2004ej,Datta:2004jm}.

Making use of the flavor topology of the decay amplitudes and the approximate
flavor SU(3) symmetry, one can describe those decay modes in terms of
several independent quark flavor flow diagrams, such as tree diagram ($\mathcal{T}$),
color-suppressed tree diagram ($\mathcal{C}$), QCD penguin diagram
($\mathcal{P}$), electroweak penguin diagram ($\mathcal{P}_{\tmop{EW}}$)
and color suppressed electroweak penguin diagram ($\mathcal{P}_{EW}^C$)
etc. It then follows from the hierarchies of the Wilson coefficients and the CKM
matrix elements that the $B \rightarrow \pi \pi$ modes are $\mathcal{T}$ dominant
while $B \rightarrow \pi K$ modes are $\mathcal{P}$
dominant. Since $\mathcal{C}$ is color suppressed, one expects that the
hierarchical structures among those decays should be
\begin{align}
2\tmop{Br} ( \pi^0 \pi^0 )&\ll \tmop{Br} ( \pi^+ \pi^- ) \approx 2\tmop{Br} ( \pi^- \pi^0 ) ,
\intertext{and}
\tmop{Br}
( \pi^+ K^- ) &\simeq \tmop{Br} ( \pi^- \bar{K}^0 ) \simeq 2 \tmop{Br} ( \pi^0
\bar{K}^0 ) \simeq 2 \tmop{Br} ( \pi^0 K^- ) ,
\end{align}
respectively.

Note that the above relations follow from a purely short distance
diagrammatic description which could be misleading in the presence of large
final state interactions (FSIs) \cite{Wolfenstein:1995zy,Neubert:1997wb,Gerard:1997kv}.
At present, they are not
favored by the  experiments.  The current world average data \cite{ICHEP04data,hfag} listed in Tab.\ref{data},
show big  enhancements of $\tmop{Br} ( \pi^0 \pi^0 )$ and
$Br(\pi^-\pi^0)$ relative to $Br( \pi^+ \pi^- )$ and suppression of $\tmop{Br} (
\pi^+ K^- )$ relative to $ 2 \tmop{Br} ( \pi^0 \bar{K}^0 ) $ and $\tmop{Br} (
\pi^- \bar{K}^0 )$.
\begin{table}[htb]
\begin{ruledtabular}
\begin{tabular}{llll}
%  \hline
  modes & $B r$($\times 10^{- 6}$) & $a_{\tmop{CP}}$ & $S$\\
  \hline
  $\pi^+ \pi^-$ & $4.6 \pm 0.4$ & $0.37 \pm 0.11$ & $- 0.61 \pm 0.14$\\
  \hline
  $\pi^0 \pi^0$ & $1.51 \pm 0.28$ & $0.28 \pm 0.39$ & \\
  \hline
  $\pi^- \pi^0$ & $5.5 \pm 0.6$ & $- 0.02 \pm 0.07$ & \\
  \hline
  $\pi^+ K^-$ & $18.2 \pm 0.8$ & $- 0.11 \pm 0.02$ & \\
  \hline
  $\pi^0 \bar{K}^0$($K_S$) & $11.5 \pm 1.0$ & $- 0.09 \pm 0.14$ & $( + 0.34
  \pm 0.28 )$\\
  \hline
  $\pi^- \bar{K}^0$ & $24.1 \pm 1.3$ & $- 0.02 \pm 0.034$ & \\
  \hline
  $\pi^0 K^-$ & $12.1 \pm 0.8$ & $0.04 \pm 0.04$ & \\
  \hline
  $K^+ K^-$ &  &  & \\
  \hline
  $K^0 \bar{K}^0$ & $1.19^{+0.42}_{-0.37}$ &  & \\
  \hline
  $K^- \bar{K}^0$ & $<2.4(1.45^{+0.53}_{-0.46})$ &  & \\
%  \hline
\end{tabular}
\end{ruledtabular}
\caption{The latest world averaged data of Charmless B decays\cite{ICHEP04data,hfag}.}
\label{data}
\end{table}
The numerical values of
these relative  ratios  are given by \cite{Buras:2004th}
\begin{eqnarray}
  R_{+ -} & = & 2 \frac{\tmop{Br} ( \pi^- \pi^0 )}{\tmop{Br} ( \pi^+ \pi^- )}
  \cdot \frac{\tau_{B^0}}{\tau_{B^+}} = 2.2 \pm 0.31 , \nonumber\\
  R_{00} & = & 2 \frac{\tmop{Br} ( \pi^0 \pi^0 )}{\tmop{Br} ( \pi^+ \pi^- )} =
  0.67 \pm 0.14 , \label{Rpipi}
\end{eqnarray}
and also
\begin{eqnarray}
  R_n & = & \frac{\tmop{Br} ( \pi^+ K^- )}{2 \tmop{Br} ( \pi^0 \bar{K}^0 )} =
  0.79 \pm 0.08 ,\nonumber\\
  %R_c & = & \frac{2 \tmop{Br} ( \pi^0 K^- )}{\tmop{Br} ( \pi^- \bar{K}^0 )} =
  %1.0 \pm 0.08 \nonumber\\
  R & = & \frac{\tmop{Br} ( \pi^+ K^- )}{\tmop{Br} ( \pi^- \bar{K}^0 )} \cdot
  \frac{\tau_{B^+}}{\tau_{B^0}} = 0.82 \pm 0.06 ,
\nonumber\\
   R_2 & = & \frac{\tmop{Br} ( \pi^+ K^- )}{2\tmop{Br} ( \pi^0 K^- )} \cdot
  \frac{\tau_{B^+}}{\tau_{B^0}}
         = 0.81 \pm 0.06 .
\end{eqnarray}
The above five ratios characterize the puzzling patterns of the latest data and
may provide insights on the strong dynamics of heavy quark decays or
possible new physics beyond the Standard Model (SM).

% R00 puzzle
The large value of $R_{00}$ forces the $\mathcal{C}$ to be large, which is
a challenge to theory. Various ways to explain large $R_{00}$ with reasonable
values of $\mathcal{C}/\mathcal{T}$ involve an enhanced $W$-exchange diagram
$(\mathcal{E})$, a large QCD penguin contribution corresponds to $u$-quark loop or
large final state interactions (FSIs) which involves $D D_{(s)}$ intermediate
states and quasi-elastic mixing between $\pi^+\pi^-$ and $\pi^0\pi^0$ modes
\cite{Cheng:2004ru}. The recent SCET calculations also supported a large 
 $\mathcal{C}/\mathcal{T}$ \cite{Bauer:2004dg}. 
%
%Note that in $\pi\pi$ and $\pi K$ modes the flavor
%topological structures are different, those effects can be in principle
%distinguished by a separate study of these two sets.  As FSIs are flavor blind,
%one expects similar values of $\mathcal{C}/\mathcal{T}$ in all $\pi\pi$ and $\pi
%K$ modes while in the case of large $\mathcal{E}$ they should be different.
%
Note that the $\pi\pi$ and $\pi K$ modes differ in flavor topological structure while
FSIs are flavor blind, the two kind of effects can in principle be distinguished by
a separated study of these two sets. FSI will lead to large $\mathcal{C}$ in all
decays modes. Furthermore, it should enhance $\mathcal{P}_{EW}^C$ relative to
$\mathcal{P}_{EW}$ in a similar manner. 

% Rn puzzle
In the $\pi K$ modes, it is well known that the suppression of $R_n$ is more relevant
to the electro-weak penguin sector, as in $\pi K$ modes $\mathcal{T}$ and
$\mathcal{C}$ are greatly suppressed by small CKM matrix elements,
%their effects
%are less important than that from electro-weak penguin diagrams
%$\mathcal{P}_{EW}$ and $\mathcal{P}_{EW}^{C}$.
In the SM, from the isospin
structure of the effective Hamiltonian, the ratio between electroweak penguin
and tree diagrams are fixed through
\cite{Neubert:1998re,Grossman:1999av,Gronau:2003kj,He:1998rq,Wu:2002nz}
\begin{eqnarray}
  R^{SM}_{EW}=\frac{P_{\tmop{EW}} + P_{\tmop{EW}}^C}{T + C} & =  & - \frac{3}{2} \cdot
  \frac{C_9 + C_{10}}{C_1 + C_2} = ( 1.35 \pm 0.12 ) \times 10^{- 2} ,
  \label{pewSM}
\end{eqnarray}
for $\pi\pi$ modes. Where $T$, $C$, $P_{EW}$ and $P_{EW}^C$ are diagrams with
CKM matrix elements factorized out which will be discussed in detail bellow. 
$C_i$s stand for the short distance Wilson
coefficients at the scale of $\mu \simeq m_b$. This relation is free from
hadronic uncertainties and survives under elastic FSIs and inelastic FSIs
through low isospin states such as $B\to DD_{s}\to \pi\pi(K)$.  It also predicts
the direct CP violation in $\pi^-\pi^0$ modes to be vanishing. Using flavor
SU(3) symmetry, this relation also holds for $\pi K$ modes. Thus it can directly
confront the experiments and allows us to explore the new physics in hadronic
charmless $B$ decays.
It is of interest that the charmless $B$ decay data indeed imply
the violation of Eq.(\ref{pewSM}). The possibility of larger
isospin $I=2(3/2)$ amplitudes violating Eq.(\ref{pewSM}) in
$\pi\pi(\pi K)$ modes was found in Ref.\cite{Zhou:2000hg} and
recently discussed
in Refs.\cite{Yoshikawa:2003hb,Atwood:2003tg,Buras:2003dj,Buras:2004ub,Buras:2004th,%
  Mishima:2004um,Nandi:2004dx,Wu:2004xx} with updated data. In a recent
analysis, an enhancement of a factor of two was obtained through a direct global
$\chi^2$ analysis using flavor SU(3) symmetry\cite{Wu:2004xx}.

Although it is too early to draw any robust conclusion, it
motivates us to take a closer look at  the electroweak penguin
sector within and beyond the SM. Note that in these analyses on
large $P_{EW}$, the effects of $P_{EW}^C$ are often assumed to be
small, which is conceptually not appropriate as $P^C_{EW}$ is
directly involved in Eq.(\ref{pewSM}). Furthermore, it provides a
cancellation to the low isospin $I=0(1/2)$ part of $P_{EW}$. Thus
its effects could be significant.

% R puzzle
The suppression of $R$ may require significant contributions from
subleading diagrams such as annihilation diagram $\mathcal{A}$ or
color-suppressed electro-weak penguin diagram $\mathcal{P}_{EW}^C$. Considering
the fact that $\mathcal{A}$ contributes to $\pi^{-}\bar{K}^0$ and $\pi^0 K^-$
in the same way, namely they have the same $\mathcal{A}-\mathcal{P}$
interference, one expects that an enhancement of $\mathcal{A}$ with appropriate
strong phase can suppress simultaneously $R$ and $R_2$ while
their effects would cancel in some extent in their ratio.
The current data give
\begin{align}
R_3=\frac{2Br(\pi^0 K^-)}{Br(\pi^{-}\bar{K}^0)}=1.0\pm 0.08 ,
\end{align}
which agrees well with this conjecture. However an enhancement of
$\mathcal{A}$ will lead to significant consequences to $KK$ modes,
especially $K^-\bar{K}^0$, as it is not suppressed by CKM matrix
element like in the $\pi K$ modes. It is expected that a strong
constraint on $\mathcal{A}$ will be found once this decay mode is
experimentally observed.

There already exists a number of global $\chi^2$ analyses on $B
\rightarrow \pi \pi, \pi K$ and $KK$ systems based on flavor SU(3) symmetry
\cite{Zhou:2000hg,He:2000ys,Chiang:2003pm,Bona:2005vz,Chiang:2004nm,Charles:2004jd,He:2004ck}.
But to trace back the origins of the above mentioned $\pi\pi$ and $\pi K$ puzzles,
$separate$ $\chi^2$ analysis are urgently needed.
% But
%in most of the previous analysises, there are no seperate discussions
%on $\pi \pi$ and $\pi K$ modes, which is important for tracing back
%the origins of the these puzzles.
Furthermore, the SU(3) breaking scheme dependences are not carefully examined in
the previous analyses, which may lead to different results in the literature.
Finally, the contributions from subleading diagrams such as
$\mathcal{P}^C_{EW}$, $\mathcal{E}$, $\mathcal{A}$ and penguin induced annihilation
diagram $\mathcal{P}_{A}$ which could play important
roles in understanding the current data are often neglected.

%One of the advatages of the diagramtic
%decompsition is that those decays modes can form  closed subsets which allows
%independent extractions of some of the decay amplitudes without inferering
%with other decay modes.
The purpose of the present paper is to make an up to the date investigation on \
charmless $B$ decays,  following  a strategy that first applying
$\chi^2$ analysis on $\pi \pi$, $\pi K$ and $K K$ modes separately, then
connecting them through flavor $\tmop{SU} ( 3 )$ symmetry and discuss
the SU(3) breaking scheme dependency.  After obtaining reasonable
values of the dominant amplitudes, we then discuss their implications
to $K K$modes with subleading diagrams such as
$\mathcal{P}_{\tmop{EW}}^C,\mathcal{E},\mathcal{A}$ $\mathcal{P}_A$
etc.

Our main observations are
\begin{itemize}
\item Independent fits on $\pi \pi$ and $\pi K$ modes without SU(3) symmetry
  both favor a large ratio between color-suppressed tree ($\mathcal{C} )$and
  tree ($\mathcal{T} )$ diagram, which disfavors the explanation of large
  nonfactorizable $W$-exchange diagrams ($\mathcal{E} )$.  The extracted QCD
  penguin diagrams from $\pi\pi$, $\pi K$ and $KK$ show a clear signal of SU(3)
  breaking and the breaking pattern is in agreement with naive factorization.

\item Global fits for $\pi \pi, \pi K$ and $ K K$ modes show good determinations
  of weak phase $\gamma$ in agreement with the standard model and prefer a
  larger electro-weak penguin $( \mathcal{P}_{\tmop{EW}} )$ relative to
  $\mathcal{T} + \mathcal{C}$ with a large strong phase when $P_{EW}^C$ is
  neglected. The results are found stable among various SU(3) breaking schemes.
  The current data favor a SU(3) breaking scheme in which all the amplitudes
  for $\pi K$ are greater by a factor of $f_K/f_\pi$ motivated from
  factorization.

\item An enhancement of $\mathcal{P}_{\tmop{EW}}^C$ with destructive
  interference to $P_{\tmop{EW}}$ provides a alternative explanation to the
  small $R_n$ within the SM. The suppression of $R$ can be partially explained
  by an enhanced annihilation diagram $\mathcal{A}$. The $\mathcal{P}_{A}$
  provides a source of SU(3) breaking in strong phases.

\item The subleading diagrams may lead to significant CP
violations in $KK$ modes. For typical value of $\mathcal{A}$ and
$\mathcal{P}$, the direct CP violation in $K^-\bar{K}^0$ can reach
$\sim 0.4$.
\end{itemize}

This paper is organized as follows. In section
\ref{sec:paremeterization}, the basic formulas for diagrammatic
decomposition are presented.  In \ref{noSU3}, we extract the
parameters such as weak phase $\gamma$ and the decay amplitudes
from $\pi \pi$, $\pi K$ and $KK$ modes separately. In section
\ref{SU3}, we combine them in three different scenarios of SU(3)
symmetry. One is that all the amplitudes in $\pi K$ modes are
rescaled by a factor of $f_K / f_{\pi}$ motivated from the native
factorization. An other one is that only the tree diagrams are
rescaled by this factor while the rest of the amplitudes remain
the same in SU(3) limit. The last one is the strict SU(3) limit.
In section \ref{nonfac}. We consider the contributions from
various subleading diagrams and extract their typical values. In
section \ref{KKmodes}. The implications to the $K K$ modes are
discussed. The possibility of finding large direct CP violations
is indicated. We conclude in section \ref{summary}.

\section{Diagrammatic description}\label{sec:paremeterization}
We use the following definitions for branching ratios and direct CP violations
\begin{eqnarray}
  \tmop{Br} & = & \frac{1}{2} \tau ( | \bar{\mathcal{A}} |^2 + |\mathcal{A}|^2 ) ,
\nonumber\\
  a_{\tmop{CP}} & = & \frac{| \bar{\mathcal{A}} |^2 - |\mathcal{A}|^2}
  { | \bar{\mathcal{A}} |^2+ |\mathcal{A}|^2 } ,
\end{eqnarray}
where $\mathcal{A}(\bar{\mathcal{A}})$ stands for $B^0(\bar{B})$ or
$B^+(B^-)$ decay amplitude. $\tau$ is a phase space factor, $\tau = 1$
for neutral final states and $\tau = \tau_{B^+} / \tau_{B^0} = 1.086
\pm 0.017$ for charged final states. The mixing induced CP violation
parameters  $S$ and $C$ are introduced  through the time-dependent decay rate
difference
\begin{eqnarray}
  a_{\tmop{CP}} ( t ) & = & \frac{\Gamma ( \bar{B}^0 \rightarrow f_{\tmop{CP}}
  ) - \Gamma ( B^0 \rightarrow f_{\tmop{CP}} )}{\Gamma ( \bar{B}^0 \rightarrow
  f_{\tmop{CP}} ) + \Gamma ( B^0 \rightarrow f_{\tmop{CP}} )} \nonumber\\
  & = & S \cdot \sin ( \Delta m_B \cdot t ) - C \cdot \cos ( \Delta m_B \cdot
  t ) ,
\end{eqnarray}
with $f_{CP}$ denoting final states with definite CP parities. $\Delta m_B$
is the neutral $B$ meson mass difference. The two parameters can be written as
\begin{eqnarray}
  S = \frac{2 \tmop{Im} \lambda}{1 + | \lambda |^2} & , & C = \frac{1 - |
  \lambda |^2}{1 + | \lambda |^2} = - a_{\tmop{CP}}  ,
\end{eqnarray}
with
\begin{align}
\lambda = e^{- 2 i \beta}  \frac{\bar{\mathcal{A}}}{\mathcal{A}} ,
\end{align}
in the SM.

Using the phase definitions of $B^-=(-\bar{u}b), \bar{B}^0=(\bar{d}b)$,
$K^-=(-\bar{u}s), \bar{K}^0=(\bar{d}s)$ and
$\pi^+=(u\bar{d}), \pi^0=(d\bar{d}-u\bar{u})/2, \pi^-=(-\bar{u}d)$,
one arrives at the following diagrammatic decompositions for
$\pi \pi$ modes \cite{Gronau:1994rj,Gronau:1995hn,Gronau:1995ng}

\begin{eqnarray}\label{diagram-pipi}
  \bar{\mathcal{A}} ( \pi^+ \pi^- ) & = & -
  (\mathcal{T}+\mathcal{E}+\mathcal{P}+\mathcal{P}_A + \frac{2}{3}
  \mathcal{P}^C_{\tmop{EW}} ) ,
  \nonumber\\
  \bar{\mathcal{A}} ( \pi^0 \pi^0 ) & = & - \frac{1}{\sqrt{2}}
  (\mathcal{C}-\mathcal{E}-\mathcal{P}-\mathcal{P}_A +\mathcal{P}_{\tmop{EW}}
  + \frac{1}{3} \mathcal{P}^C_{\tmop{EW}} ) ,
  \nonumber\\
  \bar{\mathcal{A}} ( \pi^0 \pi^- ) & = & - \frac{1}{\sqrt{2}}
  (\mathcal{T}+\mathcal{C}+\mathcal{P}_{\tmop{EW}} +\mathcal{P}^C_{\tmop{EW}}
  ) .
\end{eqnarray}
Similarly, the $\pi K$ modes are given by
\begin{eqnarray}\label{diagram-piK}
  \bar{\mathcal{A}} ( \pi^+ K^- ) & = & - (\mathcal{T}'+\mathcal{P}'+
  \frac{2}{3} \mathcal{P}^{C'}_{\tmop{EW}} ) ,
  \nonumber\\
  \bar{\mathcal{A}} ( \pi^0 \overline{K^{}}^0 ) & = & - \frac{1}{\sqrt{2}}
  (\mathcal{C}'-\mathcal{P}'+\mathcal{P}'_{\tmop{EW}} + \frac{1}{3}
  \mathcal{P}^{C'}_{\tmop{EW}} ) ,
  \nonumber\\
  \bar{\mathcal{A}} ( \pi^- \overline{K^{}}^0 ) & = & \mathcal{A}'+\mathcal{P}'-
  \frac{1}{3} \mathcal{P}^{C'}_{\tmop{EW}} ,
  \nonumber\\
  \bar{\mathcal{A}} ( \pi^0 K^- ) & = & - \frac{1}{\sqrt{2}}
  (\mathcal{T}'+\mathcal{C}'+\mathcal{A}'+\mathcal{P}'+\mathcal{P}'_{\tmop{EW}}
  + \frac{2}{3} \mathcal{P}^{C'}_{\tmop{EW}} ) .
\end{eqnarray}
The amplitudes for $\pi K$ modes are marked by a prime, which equal to the
unprimed ones for $\pi\pi$ modes in flavor SU(3) limit.
The $K K$ modes are given by
\begin{eqnarray}\label{diagram-KK}
  \bar{\mathcal{A}} ( K^+ K^- ) & = & - (\mathcal{E}''+\mathcal{P}''_A ) ,
  \nonumber\\
  \bar{\mathcal{A}} ( K^0 \bar{K}^0 ) & = & \mathcal{P}''- \frac{1}{3}
  \mathcal{P}^{C''}_{\tmop{EW}} +\mathcal{P}''_A ,
  \nonumber\\
  \bar{\mathcal{A}} ( K^- \bar{K}^0 ) & = & \mathcal{P}''- \frac{1}{3}
  \mathcal{P}^{C''}_{\tmop{EW}} +\mathcal{A}'' ,
\end{eqnarray}
where the subleading diagrams such as color-suppressed
electro-weak penguin ($\mathcal{P}_{EW}^{C}$), $W-$ exchange diagram
($\mathcal{E}$), annihilation diagram $\mathcal{A}$ and
penguin-induced annihilation diagram ($\mathcal{P}_{A}$) are included.
In the above formulas, the penguin exchange diagram ( $\mathcal{P}_{E}$ ) are absorbed into
penguin diagrams and the electroweak and color suppressed electroweak
penguin exchange diagrams are neglected.

We start with independent analyzes on $\pi \pi$, $\pi K$ and $KK$
modes.  In the first step, all subleading diagrams such as
$\mathcal{P}_{\tmop{EW}}^C$, $\mathcal{E}$, $\mathcal{P}_A$ and
$\mathcal{A}$ are switched off for the reason of simplicity. Their
effects will be investigated in detail in section \ref{SU3} and
\ref{nonfac}. To consistently include the experimental errors of
the data, the $\chi^2$ method is adopted for extracting the decay
amplitudes.  The definition of $\chi^2$ reads
\begin{align}
\chi^2=\sum_i \left(
  \frac{f^{theo}(\alpha_j)_i-f^{exp}_i}{\sigma_i}
\right)^2 ,
\end{align}
where $f^{theo}_i$ are the theoretical values of observables $f_i( i=1,m)$ and
$\alpha_j (j=1,n)$ are the to-be-determined parameters.  $f^{exp}_i$ and
$\sigma_i$ are the experimental central values and errors. The best-fit of the
parameters correspond to the minimum of the $\chi^2$ function which satisfies a
$\chi^2$ distribution with degree-of-freedom(d.o.f) $m-n$.  For the experimental
data we take the values listed in Tab.\ref{data} which are the weighted average
of CLEO, Babar and Belle collaboration results\cite{hfag}.  Other major
parameters used in the fits involve the CKM matrix element of $V_{ub}$
\cite{Aubert:2003zd,Athar:2003yg}and $V_{cb}$\cite{Hashimoto:2001nb}.
In the numerical calculations we take the following values \cite{hfag}

\begin{align}
V_{cb}=0.04\pm 0.02, \qquad V_{ub}=(3.9\pm 0.68)\times 10^{-3} ,
\end{align}
 and the SM value of \cite{Aubert:2004zt,Abe:2004mz}
\begin{align}
\sin2\beta=0.73\pm 0.037.
\end{align}
All the $Br$s are written in units of $10^{-6}$, and the angles are in gradient
and arranged in the range $(-\pi,+\pi)$.

%s2
%
%%%%%%%%%%%%%%%%%%%%%%%%%%%%%%%%%%%%%%%%%%%%%%%%%
%
\section{analysis without flavor SU(3) symmetry }\label{noSU3}
%
%%%%%%%%%%%%%%%%%%%%%%%%%%%%%%%%%%%%%%%%%%%%%%%%%

%%%%%%%%%%%%%%%%%%%%%%%%%%%%%%%%%%
\subsection{$\pi \pi$ modes}
%%%%%%%%%%%%%%%%%%%%%%%%%%%%%%%%%%
The hierarchies in the decay amplitudes are controlled by both the Wilson
coefficients and the CKM matrix elements. As the sizes of the CKM matrix
elements are better known, it is helpful to factorize them out so that all the
hadronic amplitudes in $\pi\pi$, $\pi K$ and $K K$ etc have the same
hierarchical structure.  Thus we shall use the following parameterizations for
$\pi \pi$ modes
\begin{eqnarray}
  \bar{\mathcal{A}} ( \pi^+ \pi^- ) & = & - \left[ \lambda_u ( T + E
  -P-P_A - \frac{2}{3} P^C_{\tmop{EW}} ) -
  \lambda_c (P+P_A + \frac{2}{3} P^C_{\tmop{EW}}
  ) \right] , \nonumber\\
  \bar{\mathcal{A}} ( \pi^0 \pi^0 ) & = & - \frac{1}{\sqrt{2}} \left[
  \lambda_u ( C - E + P + P_A - P_{\tmop{EW}} - \frac{1}{3} P^C_{\tmop{EW}} )
  - \lambda_c ( - P - P_A + P_{\tmop{EW}} + \frac{1}{3} P^C_{\tmop{EW}} )
  \right] , \nonumber\\
  \bar{\mathcal{A}} ( \pi^0 \pi^- ) & = & - \frac{1}{\sqrt{2}} \left[
  \lambda_u ( T + C - P_{\tmop{EW}} - P^C_{\tmop{EW}} ) - \lambda_c (
  P_{\tmop{EW}} + P^C_{\tmop{EW}} ) \right] ,
\end{eqnarray}
with $\lambda_u = V_{\tmop{ub}} V_{\tmop{ud}}^{\ast} = A \lambda^3 ( \rho - i
\eta ) ( 1 - \lambda^2 / 2 )$, and $\lambda_c = V_{\tmop{cb}}
V_{\tmop{cd}}^{\ast} = - A \lambda^3$.
% ptu
Throughout  the present paper, we shall assume the $t$-quark dominance
in penguin type diagrams. In general a penguin diagram can be written as
\begin{align}
\mathcal{P}
=\lambda_u P_u+\lambda_c P_c+\lambda_t P_t
=-\lambda_u P_{tu}-\lambda_c P_{tc} ,
\end{align}
where $P_{tu}=P_t-P_u$ and $P_{tc}=P_t-P_c$.
The $t-$ quark dominance in the Wilson coefficient then leads to
$P_t \gg P_c \gg P_u$ and
\begin{align}
P_{tu}\simeq P_{tc}\simeq P_t\equiv P.
\end{align}
Note that in the presence of large charming penguin
\cite{Ciuchini:1997hb,Ciuchini:2001gv,Isola:2001ar,Isola:2001bn,Bauer:2004dg},
$P_{tc}$ could  differ from $P$. This effect can be effective
absorbed into inelastic final state interactions(FSIs) and
will not be discussed in detail here.

From the isospin structure of the low energy effective Hamiltonian, the
sum $T+C$ and $P_{EW}+P^C_{EW}$ have both isospin $I=2$. It is then helpful to
define
\begin{align}
\hat{T}=T+C, \quad \hat{P}_{EW}=P_{EW}+P^C_{EW} .
\end{align}
The ratio $R^{SM}_{EW}$ is just the ratio of the isospin $I=2$ part between
electroweak penguin and tree diagrams.

In the naive factorization approach \cite{Ali:1998nh,Ali:1998eb},
these amplitudes have the following typical values
\begin{align}\label{naiveFac}
T &=0.9\sim 1.1, &C &= -0.33\sim 0.25 ,
\nonumber\\
P &\simeq 0.1, &P_{EW}&=0.013\sim0.015 ,
\nonumber\\
P^C_{EW}&=-0.0023\sim 0.003.  &&
\end{align}
All the amplitudes are almost real. The ranges of the parameters
correspond to the effective number of color $N_C$ ranging from 2
to infinity.  In the factorization approach, the rescaled
amplitudes satisfy a hierarchy of
\begin{align}
|T| \gg |P| \gg |P_{EW}|,|P^C_{EW}| ,
\end{align}
which holds for all primed and double-primed amplitudes in 
$\pi K$ and $K K$ modes.

Including the time dependent CP asymmetry, the $\pi\pi$ modes provides seven
data points.  A direct fit to the data gives the following best fits
corresponding to a local minimum of $\chi^2$ function.
\begin{eqnarray}\label{pipiFit}
  \abs{T} = 0.53^{+ 0.036}_{- 0.031} & , & \abs{C} = 0.42^{+ 0.081}_{- 0.11},
  \nonumber\\
  \delta_C = - 0.84^{+ 0.57}_{- 0.41} & , & \abs{P} = 0.099^{+ 0.038}_{- 0.045},
  \nonumber\\
  \delta_P = - 0.55^{+ 0.27}_{- 0.73} & , & \gamma = 1.1^{+ 0.26}_{- 0.29},
\end{eqnarray}
with $\chi^2_{\min}/d.o.f = 0.17/1$, where $P_{EW}$  is fixed relative
to $\hat{T}$ through Eq.(\ref{pewSM}).  The above result show that:

\begin{itemize}
\item The $\pi\pi$ data prefer a large $\abs{C / T}=0.8\pm 0.2$, which is in
  contradiction with the factorization based estimation. This is not new,
  however a large relative strong phase of $\delta_C=-0.84^{+0.57}_{-0.41}$ is
  also favored. Note that the recent SCET calculation which includes charming
  penguin effects prefers that $\tmop{Im} ( C / T )$ should be vanishing at
  leading order \cite{Bauer:2004dg}.  In the present $\pi\pi$ fit the charming
  penguin amplitude is not included.  The considerable uncertainties in the
  present data us from drawing a robust conclusion on the phase of $C/T$. 
  The situation will be improved when more precise data are available in the
  near future.  The large $\abs{C}$ and its strong phase $\delta_C$ are required
  by the observed two ratios in Eq.(\ref{Rpipi}). In the following figure
  (Fig.\ref{RpipiPlot}), the dependences of the ratios with $\abs{C / T}$ and
  $\delta_C$ are given. For illustration purpose, we fix other parameters to be
  $\abs{T} = 0.53$, $\abs{P} = 0.1$, $\delta_P = - 0.55$ and $\gamma = 1.1$,
  corresponding to their best fits.

\begin{figure}[htb]
\begin{center}
\includegraphics[width=0.95\textwidth]{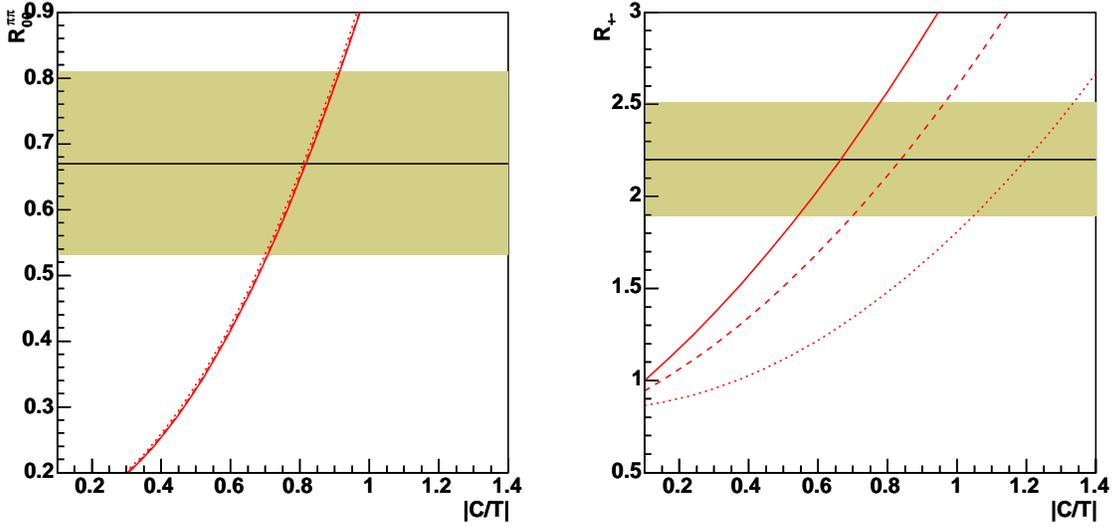}
\caption{
$R_{00}$ and $R_{+ -}$ as functions of $|C / T|$ with
different value  of $\delta_C$.The three curves correspond to $\delta_C = - 0.5, - 1.0, - 1.5$
respectively. The shadowed bands are the experimentally $1\sigma$ allowed ranges.
The other parameters are fixed at their best fitted value in Eq.(\ref{pipiFit})
}\label{RpipiPlot}.
\end{center}
\end{figure}
It follows from Fig. \ref{RpipiPlot} that both $R_{+ -}$ and $R_{00}$ prefer a
large $\abs{C / T}$ around 0.8.  There is very little dependence on $\delta_C$
for $R_{00}$. However, the large strong phase of $\delta_C \simeq - 1.0$ is
required by $R_{+ -}$, namely by the interference between $T$ and $C$ in $\pi^-
\pi^0$ mode.

\item The determination of $\gamma$ is in agreement with the SM fit.  However,
  in $\pi\pi$ system, there could be multiple solutions
  \cite{Wu:2000rb,Parkhomenko:2004mn}.  The $\chi^2$ curve as function of
  $\gamma$ is give in Fig.\ref{chisqCurveSM}, which also indicates a local
  minimum of $\chi^2$ close to $\gamma=0.23$.  But the corresponding best fitted
  other parameters are $\abs{T}\simeq 0.3$, $\abs{C}\simeq 0.84$, $\delta_C
  \simeq -1.7$ and $\abs{P}\simeq 0.36$ which looks unreasonable as it favors
  $\abs{C} \gg \abs{T}$ and $\abs{T}\simeq \abs{P}$.  To get rid of the
  multi-solutions, one may  include the $\pi K$ modes via flavor $\tmop{SU} (
  3 )$ symmetry. The simplest way is to include $\pi^+ K^-$ mode only, as it was done
  in Ref.\cite{Wu:2004xx}.  The two-fold ambiguity in $\gamma $ can be easily
  lifted.

\item The value of $\abs{P}\simeq 0.1$ agrees well with naive factorization
  estimate
  in Eq.(\ref{naiveFac}) while $T$ is suppressed. The
  enhancement of $C$ and suppression of $T$ implies that there could
be a mixing between a diagram and its color-suppressed counter part.
For $\pi\pi$ modes, it may be due to large FSI through
$B\to \pi^+\pi^-(\pi^0\pi^0) \to \pi^0\pi^0(\pi^+\pi^-)$. A recent
calculation based on  one particle exchange model indeed supports this
conjecture \cite{Cheng:2004ru}. Such a mixing may also apply to $D^0\pi^0$
and $\rho^0 \pi^0$ modes.
\end{itemize}

%%%%%%%%%%%%%%%%%%%%%%%%%%%%%
\subsection{$\pi K$ modes}
\begin{figure}[htb]
\begin{center}
\includegraphics[width=0.7\textwidth]{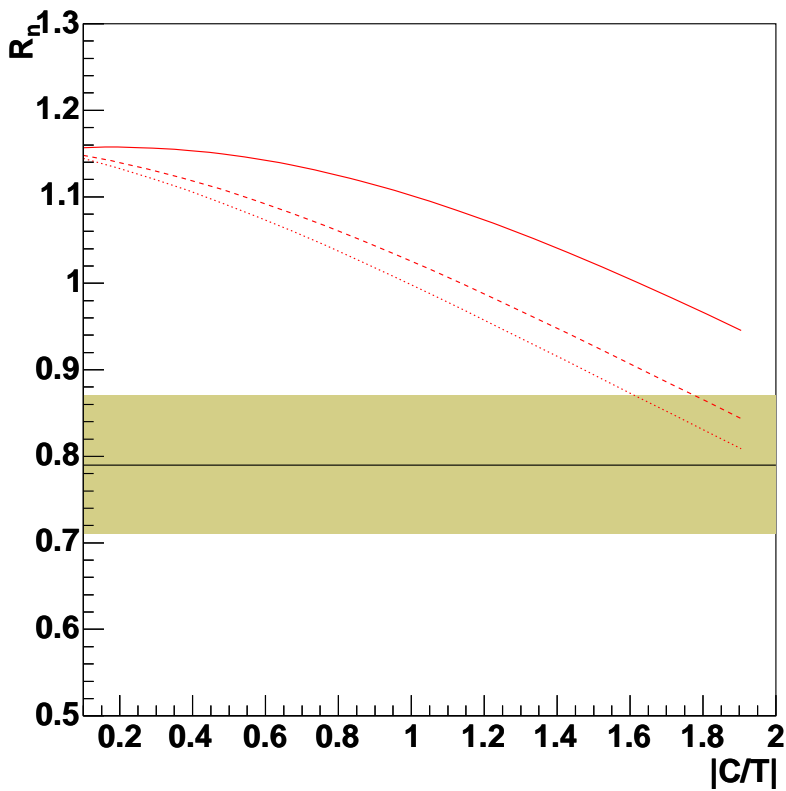}
\caption{$R_n
$as functions of $|C' / T'|$ with different value  of $\delta_{C'}$.
The three curves correspond to $\delta_{C'} = 1.0, 2.0, 3.0$ respectively.
The shadowed band is the experimentally $1\sigma$ allowed range.
}\label{c-piK.eps}
\end{center}
\end{figure}

%%%%%%%%%%%%%%%%%%%%%%%%%%%%%
The amplitudes of $\pi K$ modes are  written in a similar way

\begin{eqnarray}
  \bar{\mathcal{A}} ( \pi^+ K^- ) & = & - \left[ \lambda_u^s ( T' - P' -
  \frac{2}{3} P^{C'}_{\tmop{EW}} ) - \lambda_c^s ( P' + \frac{2}{3}
  P^{C'}_{\tmop{EW}} ) \right] ,
\nonumber\\
  \bar{\mathcal{A}} ( \pi^0 \overline{K^{}}^0 ) & = & - \frac{1}{\sqrt{2}}
  \left[ \lambda_u^s ( C' + P' - P'_{\tmop{EW}} - \frac{1}{3} P^{C'}_{\tmop{EW}}
  ) - \lambda_c^s ( - P' + P'_{\tmop{EW}} + \frac{1}{3} P^{C'}_{\tmop{EW}} )
  \right] ,
  \nonumber\\
  \bar{\mathcal{A}} ( \pi^- \overline{K^{}}^0 ) & = & \lambda_u^s ( A' - P' +
  \frac{1}{3} P^{C'}_{\tmop{EW}} ) - \lambda_c^s ( P' - \frac{1}{3}
  P^{C'}_{\tmop{EW}} ) ,
  \nonumber\\
  \bar{\mathcal{A}} ( \pi^0 K^- ) & = & - \frac{1}{\sqrt{2}} \left[
  \lambda_u^s ( T' + C' + A' - P' - P'_{\tmop{EW}} - \frac{2}{3} P^{C'}_{\tmop{EW}}
  ) - \lambda_c^s ( P' + P'_{\tmop{EW}} + \frac{2}{3} P^{C'}_{\tmop{EW}} )
  \right] , \nonumber\\
\end{eqnarray}
with $\lambda_u^s = V_{\tmop{ub}} V_{\tmop{us}}^{\ast} = A \lambda^4 ( \rho - i
\eta )$, and $\lambda_c^s = V_{\tmop{cb}} V_{\tmop{cs}}^{\ast} = A \lambda^2 ( 1
- \lambda^2 / 2 )$. Note that in the $\pi K$ modes $| \lambda_u^s |$ is much
smaller than $| \lambda_c^s |$, $| \lambda_u^s / \lambda^s_c | \simeq 0.02$.
The suppression of the tree-penguin interference and the limited accuracy of the
present data make it less effective to extract the weak phase $\gamma$ from
$\pi K$ modes at the present stage.
Thus if one considers $\pi K$ modes alone, it is more useful to take $\gamma$ as
known from the global SM fit to explore the other hadronic decay amplitudes.
Taking the SM value of $\gamma=1.08^{+0.17}_{-0.21}$ as input and also fixing
the $P_{\tmop{EW}}'$ with the SM relation of Eq.(\ref{pewSM}), one finds the
following solution
\begin{eqnarray}
  \abs{T'} = 1.54^{+ 0.61}_{- 0.38} & , & \abs{C'} = 2.7^{+ 0.61}_{- 0.7}, \nonumber\\
  \delta_{C'} = 3.1 \pm 0.11 & , & \abs{P'} = 0.12 \pm 0.0023, \\
  \delta_{P'} = - 0.2^{+ 0.07}_{- 0.12} & , &  \nonumber
\end{eqnarray}
with a $\chi^2/d.o.f = 2.49/4$. From the above result, one arrives at the following
observations
\begin{itemizedot}
\item The $\pi K$ data favor both large $T'$ and $C'$ with an even
larger ratio of $|C' / T' | = 1.75 \pm 0.7$. Although the errors
are considerably large, it is evident that a large  $|C' / T'
|\simeq \mathcal{O}(1)$ is also favored in $\pi K$ modes.  A
similar observation was made in Ref.\cite{Baek:2004rp} where no
error estimation was given.  The large $|C' / T' |$ is due to the
suppression of $R_n$ from unity, in Fig.\ref{c-piK.eps} the value
of $R_n$ as function of $\abs{C'/T'}$ is plotted, one sees that in
general, a large $\abs{C'/T'}$ with large relative strong phase
$\delta_{C'}\simeq 2$ can lead to the reduction of the ratio
$R_n$.

\item $P'$ is well determined which is about 20 \% larger than $P$,
  in a good agreement with the factorization based estimation that
  $P'/P\simeq f_K/f_\pi\simeq 1.28$.

\item A relatively larger $\chi^2/d.o.f$ in $\pi K$ fit indicates larger
  inconsistency with the data in comparison with that for $\pi\pi$
  modes . The sources of inconsistency mainly come from $\tmop{Br} (
  \pi^+ K^- )$ and $\tmop{Br} ( \pi^0 \bar{K}^0 )$ and also $S ( \pi^0
  K_S )$. The best fit prefers a larger value of $R_n \simeq 0.9$ and
  a very small $S ( \pi^0 K_S ) \simeq - 0.02$.
\end{itemizedot}

An alternative way to achieve at a smaller  $R_n$ is to allow
$P_{\tmop{EW}}$ to be larger, which needs new physics effects
beyond the SM. Taking $P_{\tmop{EW}}$ to be free, one finds
\begin{eqnarray}
  \abs{T'} = 2.75^{+ 1.12}_{- 1.38}  & , & \abs{C'} = 1.31^{+ 0.71}_{- 0.75} ,
  \nonumber\\
  \delta_{C'} = 2.76 \pm 0.28 & , & \abs{P'} = 0.12 \pm 0.0023, \nonumber\\
  \delta_{P'} = - 0.08^{+ 0.02}_{- 0.08} & , & \abs{P_{\tmop{EW}}'} = 0.048 \pm 0.02 ,
  \nonumber\\
  \delta_{P_{\tmop{EW}}'} = 1.44^{+ 0.08}_{- 0.13} .&  &
\end{eqnarray}
with $\chi^2/d.o.f=0.4/2$.  Indeed, one sees that a large $P'_{\tmop{EW}}$ is
favored by the $\pi K$ data with $|P'_{\tmop{EW}} / ( T' + C' ) | = 0.04 \pm
0.04$. Once $P'_{\tmop{EW}}$ is increasing, the ratio of $C' / T'$ decreases.
It seems to be a promising way to restore a reasonable value of $C' / T'$.
However, it only holds for $\pi K$ modes. Furthermore, the uncertainties are too
large to prevent us to draw any robust conclusion on that.  The precisions can
be improved significantly by making use of the whole charmless $B$ decay data
connected by flavor SU(3) symmetry.

It is more difficult to explain  the suppression of $R$ in
$\pi^+K^-$ and $\pi^-\bar{K}^0$ modes, as both  $C$ and
$P_{\tmop{EW}}$  are absent. If the puzzle of $R$ has to be taken
seriously, one needs an enhancement of either  $A$  or
$P_{\tmop{EW}}^C$. This possibility will be discussed in sections
\ref{SU3} and \ref{nonfac}.

%%%%%%%%%%%%%%%%%%%%%%%%%
\subsection{$KK$ modes}
%%%%%%%%%%%%%%%%%%%%%%%%%
For the $KK$ modes, currently only $K^0\bar{K}^0$ mode has been observed.
The decay amplitude is given by
\begin{align}
 \bar{\mathcal{A}} ( K^0 \bar{K}^0 ) =  \lambda_u ( - P'' + \frac{1}{3}
  P^{C''}_{\tmop{EW}} - P''_A ) - \lambda_c ( P'' - \frac{1}{3} P^{C''}_{\tmop{EW}} +
  P''_A ) ,
\end{align}
which is dominated by QCD penguin (through $b\to d$).
Neglecting small subleading diagrams $P_{EW}^C$ and $P_A $, one can directly extract the  amplitude
of penguin from the data
\begin{align}
\abs{P''}=0.2^{+0.4}_{-0.3} ,
\end{align}
where we have taken the SM value of $\gamma$ as input. It is evident that
the QCD penguins for $\pi\pi$, $\pi K$ and $KK$ follow a pattern
\begin{align}
\abs{P}\alt \abs{P'} \alt \abs{P''} ,
\end{align}
and roughly agrees with a factorization based estimation in that
the SU(3) breaking effects are proportional to the decay constants
of the final states, namely
\begin{align}
\frac{P'}{P}\approx \frac{P''}{P'}\approx \frac{f_K}{f_\pi} .
\end{align}
Thus one finds that a separate analysis can indeed provide important information on
decay amplitudes and  test SU(3) symmetry breaking, which can not be obtained by a global
$\chi^2$ analysis.
%s3
%%%%%%%%%%%%%%%%%%%%%%%%%%%%%%%%%%%%%%%%%%%%%%%%%%
\section{analysis using flavor $\tmop{SU}$(3) symmetry}\label{SU3}
%%%%%%%%%%%%%%%%%%%%%%%%%%%%%%%%%%%%%%%%%%%%%%%%%%
%
\subsection{Fit within SM}
We are in the position now to
connect  the $\pi \pi$ , $\pi K$ and $K K$ modes through
approximate flavor SU(3) symmetry. Note that there is no reliable way
to estimate the size of the SU(3) breaking in theory.
From the factorization based approaches the SU(3) breaks in such a way that
the amplitudes in $\pi K(KK)$ modes differ from the ones in
$\pi\pi(\pi K)$ modes by a factor of  $f_K/f_\pi$,  where $f_K$
and $f_\pi$ are decay constants for $K$ and $\pi$ mesons. There
have been analysis based on different  patterns of SU(3) breaking
which could in general lead to different results. Here we would
like to consider three scenarios of SU(3) relations frequently
used in the literature:

{\bf Scenario A}) All diagrammatic amplitudes for $\pi K(KK)$ modes are
larger than that in $\pi\pi(\pi K)$ modes  by a factor $f_K / f_{\pi}$.
\begin{eqnarray}
  \frac{T'}{T} = \frac{C'}{C} = \frac{P'}{P} \cdots
=\frac{T''}{T'} = \frac{C''}{C'} = \frac{P''}{P'} \cdots
= \frac{f_K}{f_{\pi}}, &  &
\end{eqnarray}

{\bf Scenario B}) SU(3) symmetry breaks only in  tree diagrams \cite{Chiang:2003pm,Chiang:2004nm}
\begin{eqnarray}
  \frac{T'}{T}= \frac{T''}{T'} = \frac{f_K}{f_{\pi}} & , &
\frac{C'}{C} = \frac{P'}{P} \cdots
=\frac{C''}{C'} = \frac{P''}{P'} \cdots
  = 1 ,
\end{eqnarray}

{\bf Scenario C}) Exact SU(3) limit
\begin{eqnarray}
  \frac{T'}{T} = \frac{C'}{C} = \frac{P'}{P} \cdots
=\frac{T''}{T'} = \frac{C''}{C'} = \frac{P''}{P'} \cdots
= 1 .
\end{eqnarray}

As in the first step, the $P_{EW}$ is fixed  within the  SM through  Eq.(\ref{pewSM}).
Thus there are 6  parameters $T$, $C$, $\delta_C$, $P$, $\delta_P$ and $\gamma$ to be fitted by 18 data points.  The best-fitted parameters as well
as the $Br$s and $a_{CP}$s are tabulated  in Tab.\ref{SMFit}.

\begin{figure}[htb]
\begin{center}
\includegraphics[width=0.7\textwidth]{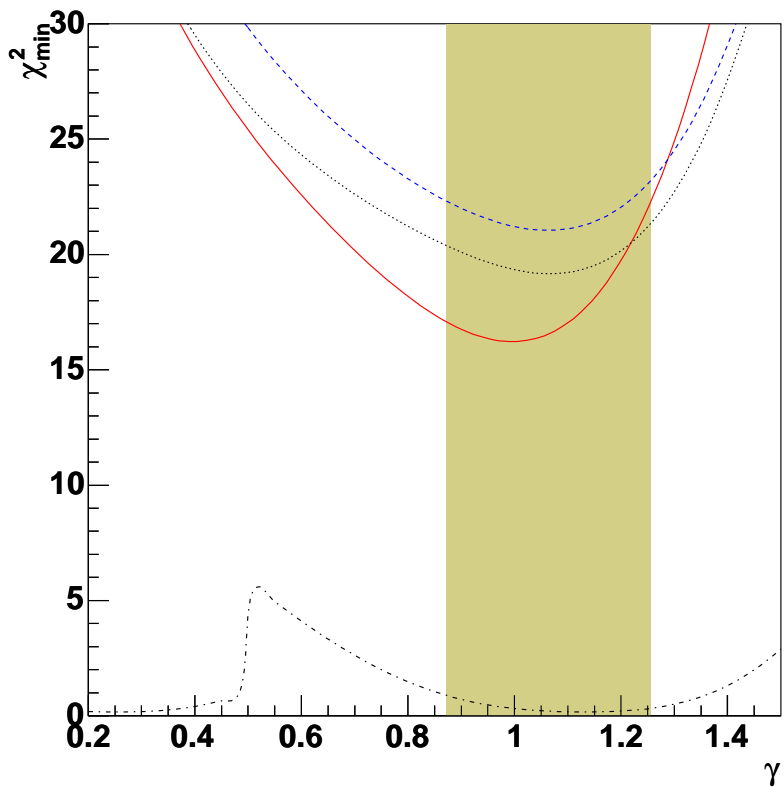}
\caption{$\chi^2_{min}$ as functions of weak phase $\gamma$.
  The three upper curves correspond to the three fit in Tab.\ref{SMFit}.  The
  solid, dashed and dotted curves correspond to scenario A, B and C
  respectively. The lower curve (dot-dashed) corresponds to the fit only to
  $\pi\pi$ modes in Eq.(\ref{pipiFit}). The shadowed band is the
  allowed range from global SM fit.  }
\label{chisqCurveSM}
\end{center}
\end{figure}

\begin{table}[htb]
\begin{center}
\begin{ruledtabular}
\begin{tabular}{llll}
                          & scenario A              & scenario B              & scenario C      \\ \hline
$\abs{T}$                 &$0.52\pm0.027$           &$0.51\pm0.033$           &$0.52\pm0.032$           \\
$\abs{C}$                 &$0.47\pm0.042$           &$0.45\pm0.053$           &$0.45\pm0.053$           \\
$\delta_C$                &$-1.1\pm0.19$            &$-1^{+0.21}_{-0.19}$     &$-1.1^{+0.21}_{-0.19}$   \\
$\abs{P}$                 &$0.094\pm0.0014$         &$0.12\pm0.0019$          &$0.12\pm0.0018$          \\
$\delta_P$                &$-0.49^{+0.089}_{-0.1}$  &$-0.45^{+0.087}_{-0.11}$ &$-0.54^{+0.11}_{-0.13}$  \\
$\abs{P_{EW}}$            &$0.011\pm0.0011$         &$0.011\pm0.0011$         &$0.011\pm0.0011$         \\
$\delta_{P_{EW}}$         &$-0.52\pm0.1$            &$-0.47\pm0.11$           &$-0.49\pm0.11$           \\
$\gamma$                  &$1^{+0.11}_{-0.13}$      &$1.1^{+0.13}_{-0.17}$    &$1.1^{+0.14}_{-0.17}$    \\
$\chi^2_{min}/d.o.f$      &16.2/12                  &19.2/12                  &21/12                       \\
$Br(\pi^+\pi^-)$          &$4.7\pm0.48$             &$4.8\pm0.62$             &$4.9\pm0.59$             \\
$a_{CP}(\pi^+\pi^-)$      &$0.27\pm0.062$           &$0.32\pm0.085$           &$0.37\pm0.096$           \\
$Br(\pi^0\pi^0)$          &$1.7\pm0.31$             &$1.8\pm0.4$              &$1.8\pm0.41$             \\
$a_{CP}(\pi^0\pi^0)$      &$0.36\pm0.11$            &$0.43\pm0.15$            &$0.38\pm0.17$            \\
$Br(\pi^-\pi^0)$          &$5.2\pm0.77$             &$5.1\pm0.85$             &$5.1\pm0.86$             \\
$a_{CP}(\pi^-\pi^0)$      &$0\pm0.01$               &$0\pm0.01$               &$0\pm0.01$               \\
$Br(\pi^+K^-)$            &$20\pm0.77$              &$20\pm0.84$              &$20\pm0.74$              \\
$a_{CP}(\pi^+K^-)$        &$-0.1\pm0.02$            &$-0.097\pm0.022$         &$-0.089\pm0.019$         \\
$Br(\pi^0\bar{K}^0)$      &$9.8\pm0.49$             &$9.9\pm0.47$             &$9.7\pm0.46$             \\
$a_{CP}(\pi^0\bar{K}^0)$  &$-0.1\pm0.035$           &$-0.076\pm0.03$          &$-0.068\pm0.032$         \\
$Br(\pi^-\bar{K}^0)$      &$22\pm0.69$              &$22\pm0.71$              &$22\pm0.68$              \\
%CP(PI-K0b)               &$-7e-19\pm1.4e-16$       &$2.5e-19\pm1.4e-16$      &$-2.8e-19\pm1.3e-16$     \\
$Br(\pi^0 K^-)$           &$12\pm0.63$              &$11\pm0.64$              &$12\pm0.57$              \\
$a_{CP}(\pi^0 K^-)$       &$0.0055\pm0.042$         &$-0.016\pm0.039$         &$-0.014\pm0.04$          \\
%Br(K+K-)                 &$0\pm0$                  &$0\pm0$                  &$0\pm0$                  \\
%CP(K+K-)                 &$0\pm0$                  &$0\pm0$                  &$0\pm0$                  \\
$Br(K^0\bar{K}^0)$        &$1.3\pm0.17$             &$2.3\pm0.35$             &$0.84\pm0.13$            \\
%CP(K0K0b)                &$6.7e-20\pm1.4e-16$      &$4.6e-20\pm1.3e-16$      &$3.5e-19\pm1.3e-16$      \\
$Br(K^-\bar{K}^0)$        &$1.3\pm0.17$             &$2.3\pm0.35$             &$0.84\pm0.13$            \\
%CP(K-K0b)                &$6.7e-20\pm1.4e-16$      &$4.6e-20\pm1.3e-16$      &$3.5e-19\pm1.3e-16$      \\
$S(\pi^+\pi^-)$           &$-0.73\pm0.13$           &$-0.76\pm0.13$           &$-0.73\pm0.14$           \\
$S(\pi^0\pi^0)$           &$0.23\pm0.27$            &$0.51\pm0.27$            &$0.52\pm0.27$            \\
$S(\pi^0 K_S)$         &$0.86\pm0.038$          &$0.84\pm0.04$           &$0.84\pm0.04$           \\
\end{tabular}
\end{ruledtabular}
\end{center}
\caption{Global fit to $\pi\pi$, $\pi K$ and $KK$ modes within SM. The three columns
corresponds to the three different SU(3) relations used in the fits.}\label{SMFit}
\end{table}
The  results show that:
\begin{itemize}
\item The differences among the three scenarios are in general not large. The weak
  phase $\gamma$ is well determined in all the cases and depends on the SU(3)
  breaking scheme very weakly. All the three scenarios give $\gamma \simeq 1.1$
  in a good agreement with the SM value with differences less than $10 \%$,
  which manifests that $\gamma$ can be reliably extracted using the diagrammatic
  approach. The $\chi^2_{min}$ curves as function of $\gamma$ are given in
  Fig.\ref{chisqCurveSM}., Comparing with the one from $\pi\pi$ fit, one finds a
  significant improvement on the precision of $\gamma$ determination. The three
  patterns lead to roughly the same $\abs{T}$ and $\abs{C}$ with $\abs{C/T}
  \simeq 0.8$. Note that for small $\abs{C}$ we find no consistent fit. For example, if
  $\abs{C}$ is fixed at $C=0.2$, a very big $\chi^2_{min}/d.o.f=44.6/12$ is obtained.
  The major difference is the value of $\abs{P}$.  The scenario B) and C) prefer a $\abs{P}$
  which is $\sim 20 \%$ larger.

\item Among the three cases, the scenario A) that all the primed (double-primed)
  amplitudes are larger than the unprimed (primed ) ones by a factor of $f_K /
  f_{\pi}$ gains the lowest $\chi^2/d.o.f=16.2/12$, which indicates a better
  consistency in comparison with the other two. The exact $SU(3)$ scenario gains
  the largest $\chi^2/d.o.f=21/12$, which clearly indicates that the flavor $SU(3)$
  symmetry in charmless $B$ decays must be a broken one.

\item The main source of the inconsistency comes from the $\tmop{Br} ( \pi^+ K^-
  )$, $\tmop{Br} ( \pi^0 \overline{K^{}}^0 )$ and $\tmop{Br} ( \pi^-
  \overline{K^{}}^0 )$. The best fits in scenario A prefer a larger $\tmop{Br} (
  \pi^+ K^- ) \simeq 20$, a small $\tmop{Br} ( \pi^0 \overline{K^{}}^0 ) \simeq
  9.8$ and a small $\tmop{Br} ( \pi^- \overline{K^{}}^0 ) \simeq 22$. Namely, within
  the current parameterization, it is not possible to arrive at the observed
  ratios $R_n$ and $R$. Thus the $\pi K$ puzzles remain.

 \item For the predictions for the $KK$ modes, the scenario A gives
$Br(K^0\bar{K}^0)=Br(K^-\bar{K}^0)=1.2$, while the other two give 1.7(scenario B)
and 0.84 (scenario C) respectively. The branching ratio of $K^+K^{-}$ is predicted
to be zero and all  the predicted direct CP violation are vanishing,
due to the lack of interferences between amplitudes.

\end{itemize}
Other possibilities of SU(3) breaking include the SU(3) breaking in strong
phases, which has been discussed in Refs.\cite{Wu:2002nz}. The current data
favor a small $SU(3)$ breaking in the strong phase of QCD penguin. This breaking
effect can significantly modify  the correlation of direct CP asymmetries between $\pi\pi$ and
$\pi K$ modes \cite{Wu:2004xx}.

\subsection{Fit with free electroweak penguin $P_{EW}$}
In the next step, we consider the possibility that $P_{\tmop{EW}}$
is free from the SM constraint. The fitting results with free
$P_{EW}$ under three scenarios are given in Tab.\ref{nonSMFit}.

\begin{table}[htb]
\begin{center}\begin{ruledtabular}
\begin{tabular}{llll}
                          & scenario A              & scenario B              & scenario C      \\ \hline
$\abs{T}$                    &$0.52\pm0.028$           &$0.51^{+0.037}_{-0.045}$ &$0.51\pm0.035$           \\
$\abs{C}$                    &$0.45\pm0.052$           &$0.44^{+0.096}_{-0.062}$ &$0.44^{+0.078}_{-0.064}$ \\
$\delta_C$                   &$-0.96^{+0.23}_{-0.21}$  &$-0.92\pm0.25$           &$-0.98\pm0.24$           \\
$\abs{P}$                    &$0.093\pm0.0015$         &$0.12\pm0.0019$          &$0.12\pm0.0019$          \\
$\delta_P$                   &$-0.52^{+0.1}_{-0.13}$   &$-0.49^{+0.1}_{-0.24}$   &$-0.59^{+0.13}_{-0.22}$  \\
$\abs{P_{EW}}$               &$0.023^{+0.0096}_{-0.011}$&$0.027\pm0.014$          &$0.029\pm0.013$          \\
$\delta_{P_{EW}}$            &$0.63^{+0.21}_{-0.41}$   &$0.7^{+0.23}_{-0.35}$    &$0.63^{+0.23}_{-0.32}$   \\
$\gamma$                     &$1^{+0.13}_{-0.18}$      &$1.1^{+0.16}_{-0.36}$    &$1^{+0.17}_{-0.26}$      \\
$\chi^2_{min}/d.o.f$         &13.2/10                  &15.9/10                  &18/10                       \\
$Br(\pi^+\pi^-)$             &$4.7\pm0.53$             &$4.8\pm0.84$             &$4.9\pm0.7$              \\
$a_{CP}(\pi^+\pi^-)$         &$0.29\pm0.077$           &$0.34\pm0.14$            &$0.39\pm0.14$            \\
$Br(\pi^0\pi^0)$             &$1.6\pm0.38$             &$1.7\pm0.66$             &$1.7\pm0.58$             \\
$a_{CP}(\pi^0\pi^0)$         &$0.14\pm0.18$            &$0.16\pm0.26$            &$0.12\pm0.26$            \\
$Br(\pi^-\pi^0)$             &$5.4\pm0.89$             &$5.3\pm1.2$              &$5.2\pm1.1$              \\
$a_{CP}(\pi^-\pi^0)$         &$-0.085\pm0.045$         &$-0.11\pm0.061$          &$-0.11\pm0.06$           \\
$Br(\pi^+K^-)$               &$20\pm0.85$              &$20\pm1.1$               &$20\pm0.84$              \\
$a_{CP}(\pi^+K^-)$           &$-0.11\pm0.024$          &$-0.1\pm0.035$           &$-0.093\pm0.027$         \\
$Br(\pi^0\bar{K}^0)$         &$11\pm1.9$               &$11\pm1.8$               &$11\pm1.9$               \\
$a_{CP}(\pi^0\bar{K}^0)$     &$-0.034\pm0.045$         &$-0.027\pm0.043$         &$-0.02\pm0.042$          \\
$Br(\pi^-\bar{K}^0)$         &$22\pm0.73$              &$22\pm0.74$              &$22\pm0.73$              \\
%CP(PI-K0b)                   &$-2e-19\pm1.3e-16$       &$-7.9e-19\pm1.3e-16$     &$2.8e-19\pm1.3e-16$      \\
$Br(\pi^0 K^-)$              &$12\pm2.2$               &$12\pm2.2$               &$12\pm2.2$               \\
$a_{CP}(\pi^0 K^-)$          &$0.033\pm0.059$          &$0.012\pm0.063$          &$0.01\pm0.057$           \\
%Br(K+K-)                     &$0\pm0$                  &$0\pm0$                  &$0\pm0$                  \\
%CP(K+K-)                     &$0\pm0$                  &$0\pm0$                  &$0\pm0$                  \\
$Br(K^0\bar{K}^0)$           &$1.3\pm0.21$             &$2.3\pm0.56$             &$0.81\pm0.17$            \\
%CP(K0K0b)                    &$-1.1e-18\pm1.4e-16$     &$-2.1e-19\pm1.3e-16$     &$-5e-19\pm1.3e-16$       \\
$Br(K^-\bar{K}^0)$           &$1.3\pm0.21$             &$2.3\pm0.56$             &$0.81\pm0.17$            \\
%CP(K-K0b)                    &$-1.1e-18\pm1.4e-16$     &$-2.1e-19\pm1.3e-16$     &$-5e-19\pm1.3e-16$       \\
$S(\pi^+\pi^-)$              &$-0.7\pm0.17$            &$-0.73\pm0.22$           &$-0.71\pm0.2$            \\
$S(\pi^0\pi^0)$              &$0.4\pm0.31$             &$0.54\pm0.41$            &$0.55\pm0.36$            \\
$S(\pi^0 K_S)$            &$0.86\pm0.039$          &$0.84\pm0.043$          &$0.84\pm0.042$          \\
\end{tabular}
\end{ruledtabular}\end{center}
\caption{The same as Tab.\ref{SMFit}, but $P_{EW}$ is taken as a free
parameter to be determined directly from the data.}
\label{nonSMFit}
\end{table}

\begin{figure}[htb]%\label{}
\begin{center}
\includegraphics[width=0.7\textwidth]{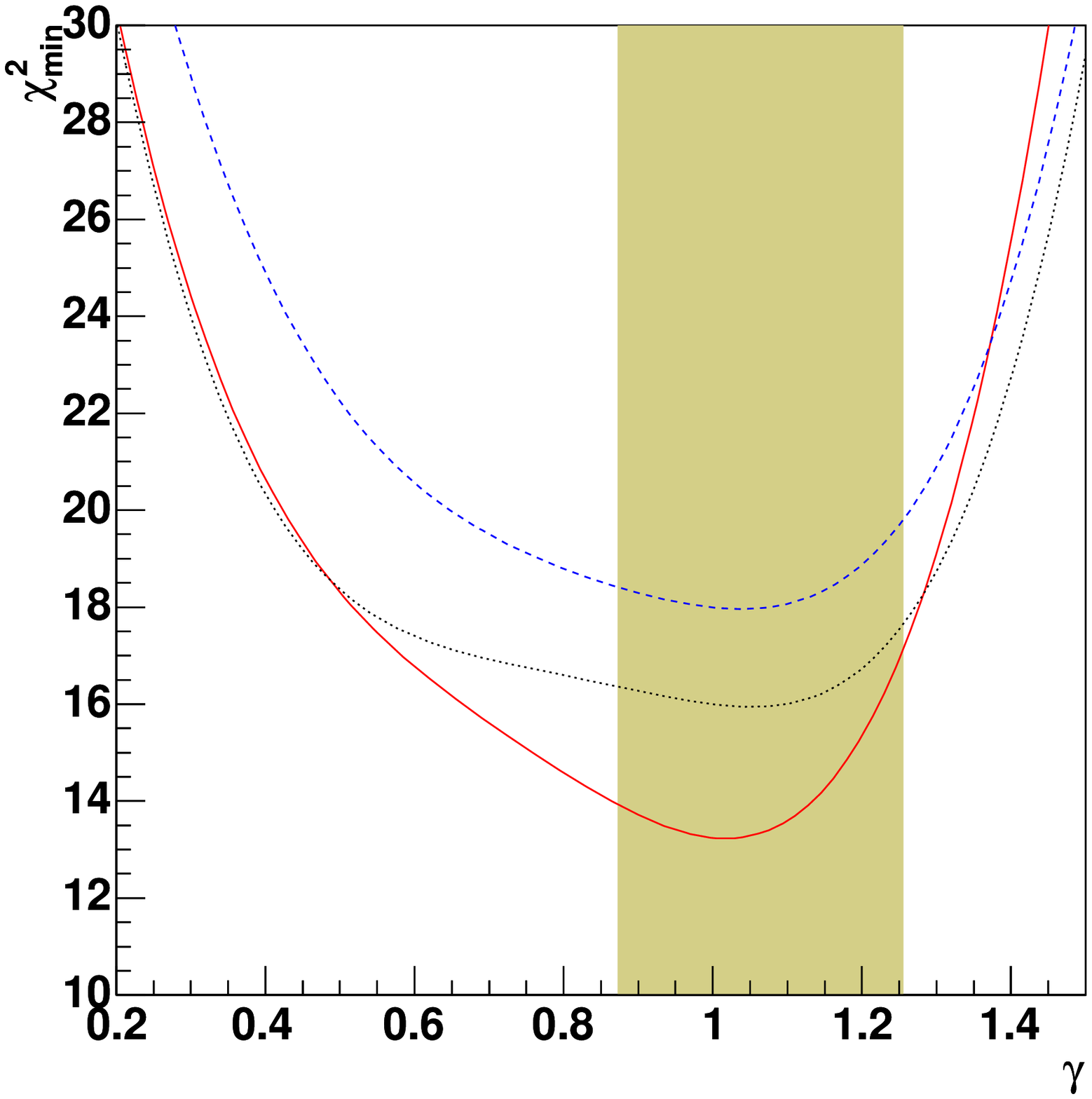}
\caption{ $\chi^2_{min}$ as functions of weak phase $\gamma$.
The three upper curves correspond to the three fit in Tab.\ref{nonSMFit}.
The solid, dashed and dotted curves correspond to scenario A, B and C
respectively.  The shadowed band is the allowed range from global SM fit.
}
\label{chisqDistr-2}
\end{center}
\end{figure}

In this case, one finds that all the fits prefer a larger value of
$\abs{P_{\tmop{EW}}}=0.23\sim 0.29$ with a large strong phase
$\delta_{P_{EW}}=0.6\sim 0.7$ relative to $\hat{T}$ as the corresponding best
fit of $\hat{T}$ has a strong phase of $-0.5\sim -0.4$. A large $P_{\tmop{EW}}$
with a large strong phase relative to $\hat{T}$ can naturally  explain the suppression of
$R_n$ and also $R_2$ \cite{Yoshikawa:2003hb,Atwood:2003tg,Buras:2003dj,Buras:2004ub,Buras:2004th,%
Mishima:2004um,Nandi:2004dx,Wu:2004xx}.
%\begin{eqnarray}
%  R_2 & = & \frac{\tmop{Br} ( \pi^+\bar{K}^-)}{2\tmop{Br} ( \pi^0 K^- )} \cdot
%  \frac{\tau_{B^+}}{\tau_{B^0}} = 0.81 \pm 0.06
%\end{eqnarray}
Naively speaking, all $\pi K$ modes are QCD penguin dominant, the
ratios $R_n$ and $R_2$ should be very close to unity. The
corrections arise from either tree type diagrams or electroweak
penguins. The former is CKM suppressed in $\pi K$ modes and is
constrained by $\pi\pi$ data. Thus an enhancement of electroweak
penguin is needed.

\begin{figure}[htb]
\begin{center}
\includegraphics[width=0.95\textwidth]{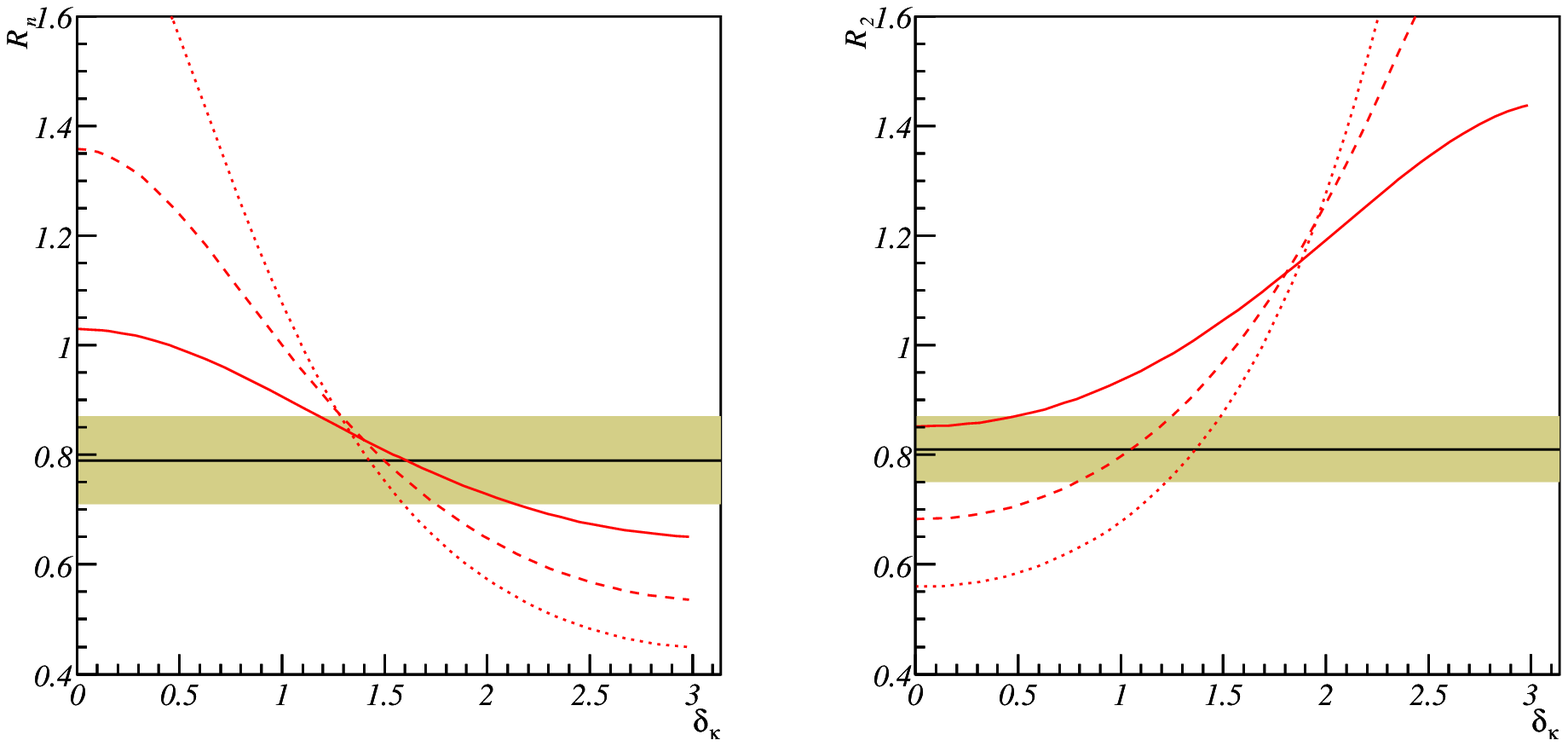}
\caption{$R_n$ and $R_2$ as function of $\delta_{\kappa}$ with different value of $|
  \kappa |$. The three curves correspond to $| \kappa | = 1.0 ( \tmop{solid} ),
  2.0 ( \tmop{dashed}, 3.0$ (dotted). The shadowed bands are the experimentally
  allowed ranges.  Other parameter are fixed at the SM best fitted values (column
  A in Tab.\ref{SMFit}). }
\label{pew}
\end{center}
\end{figure}

As the two ratios $R_n$ and $R_2$ are both $P_{\tmop{EW}}^{}$ sensitive,
they can be used as probes of electro-weak penguins.
We parameterize the deviation of SM by introducing a complex parameter $\kappa$
\begin{eqnarray}
  \frac{\hat{P}_{EW}}{\hat{T}} & = & R^{\text{SM}}_{\text{EW}} \cdot \kappa .
\end{eqnarray}
In Fig.\ref{pew}, the two ratios are plotted with different values
of $| \kappa |$ and its strong phase $\delta \kappa$. For
demonstration reasons, the other parameters are fixed at the best
fitted values in SM in the first column of Tab. \ref{SMFit}
according to the scenario A, namely
\begin{align}\label{SA}
T = 0.52, \quad C = 0.47, \quad \delta_C = - 1.1,
\quad P = 0.094 \quad \text{and} \quad \delta_P = - 0.49 .
\end{align}

The figures indicate strong dependences on $\delta \kappa$ and  $|
\kappa |$ for both ratios. In the SM case, i.e $| \kappa |$=1 and
$\delta_{\kappa} = 0$, $R_n$ is expected to be above 1.0 in
contrast with the experiment.  For $| \kappa |$=1 and large
$\delta_{\kappa} \sim 1.5$, $R_n$ is reduced but  $R_2$ is
enhanced and departs away from the allowed range of the data. Thus
to simultaneously explain both measurements, one needs a large $|
\kappa | \simeq 2.0 \sim 3.0$ and a large strong phase of
$\delta_{\kappa} \simeq 1.0 \sim 1.5$. This  observation is
confirmed by the global fits with $P_{\tmop{EW}}$ free in Tab.
\ref{nonSMFit} which gives
\begin{align}\label{eq:largePew}
\frac{P_{EW}}{\hat{T}} =
\left\{
\begin{array}{ll}
( 3.1 \pm 1.3 ) e^{i(1.02\pm 0.5)} \times 10^{- 2} &\qquad\text{(scenario A)}\\
( 4.8 \pm 1.5 ) e^{i(1.06\pm 0.53)}\times 10^{- 2} &\qquad\text{(scenario B)}\\
( 3.1 \pm 1.5 ) e^{i(1.03\pm 0.5)} \times 10^{- 2} &\qquad\text{(scenario C)} .
\end{array}
\right.
\end{align}
With $P_{EW}$ being free , the best fitted $\pi^0\bar{K}^0$ mode is found to be
$Br(\pi^0\bar{K}^0)=11\pm 1.9$ in scenario A, in a remarkable agreement with the
data. The central values of the two ratios are found to be $R_n \simeq R_2 \simeq
0.9$.

% CP(pi- pi0)
The enhanced $P_{\tmop{EW}}$ with large  strong phase will result
in different predictions for the CP asymmetries. In most decay
modes the predicted $a_{CP}$s are much smaller \cite{Wu:2004xx}.
However, the most important  one is $a_{\tmop{CP}} ( \pi^- \pi^0$)
which should be exactly zero in SM. But now it prefers a negative
value of $a_{CP}(\pi^-\pi^0)\sim - 0.1$, which is in agreement
with the current preliminary data of $a_{CP}=-0.02\pm 0.07$  and
can be examined with more precise data in the near future.

% comments on Li's work
It needs to be emphasized that the {\it large} $P_{EW}$ here means
a relative enhancement to tree type diagram $\hat{T}$, not to QCD
penguin diagram. It was claimed recently in Ref.
\cite{Charng:2004ed} that there was no clear indication of large
$P_{EW}/P$, which does not necessarily contradict with the
conclusions in the present paper.  Furthermore, using the
numerical value of $T$ and $C$ obtained in Ref.
\cite{Charng:2004ed}, we find a similar result as in
Eq.(\ref{eq:largePew}).  Note that the ratio $P_{EW}/P$ is
subjected to large theoretical uncertainties, it is better to use
the values relative to $\hat{T}$ for exploring the electro-weak
penguin and new physics as it is free from hadronic uncertainties.

%%%%%%%%%%%%%%%%%%%%%%%%%
\subsection{Effects of color suppressed electroweak penguin $P_{\tmop{EW}}^C$}
%%%%%%%%%%%%%%%%%%%%%%%%%

%We proceed to discuss the effects on subleading diagrams.
%Among all these diagrams, the color-suppressed electro-weak penguin
In the previous discussions, the color-suppressed electroweak
penguin diagram $P_{EW}^C$ is neglected. However, among all the
subleading diagrams $P_{\tmop{EW}}^C$ is the only one giving
contribution to high isospin state $I = 2 ( 3 / 2 )$ in $\pi
\pi$($\pi K )$ modes. Furthermore, it cancels the isospin
$I=0(1/2)$ components of $P_{EW}$ just like $C$ cancels that of
$T$ and directly contribute to the ratio in Eq.(\ref{pewSM}).
Since the current data indicate a sizable $C$, it is still
possible that there will be an enhancement of $P_{\tmop{EW}}^C$ as
well. In the SM, $\hat{P}_{EW}$ is fixed relative to $\hat{T}$.
However, given a large relative strong phase i.e negative
interference between $P_{\tmop{EW}}$ and $P_{\tmop{EW}}^C$, a
large value of $P_{\tmop{EW}}^C$ is possible within the SM. Note
that the $\pi^0 \bar{K}^0$ mode depends on $P_{\tmop{EW}} +
P_{\tmop{EW}}^C/ 3 = \hat{P}_{EW}/3+ 2 P_{\tmop{EW}} / 3$. Even
the first term is constrained by Eq.(\ref{pewSM}), the second term
can still
%a large $P_{\tmop{EW}}^C$ destructively interfering  with
%$P_{\tmop{EW}}$ can
enhance the decay rate of $\pi^0 \bar{K}^0$ mode without violating the SM
relation. The similar arguments also applies to $\pi^0 K^-$ modes which
depends on $P_{EW}+2 P_{EW}^C/3$.

\begin{figure}[htb]
\begin{center}
\includegraphics[width=0.95\textwidth]{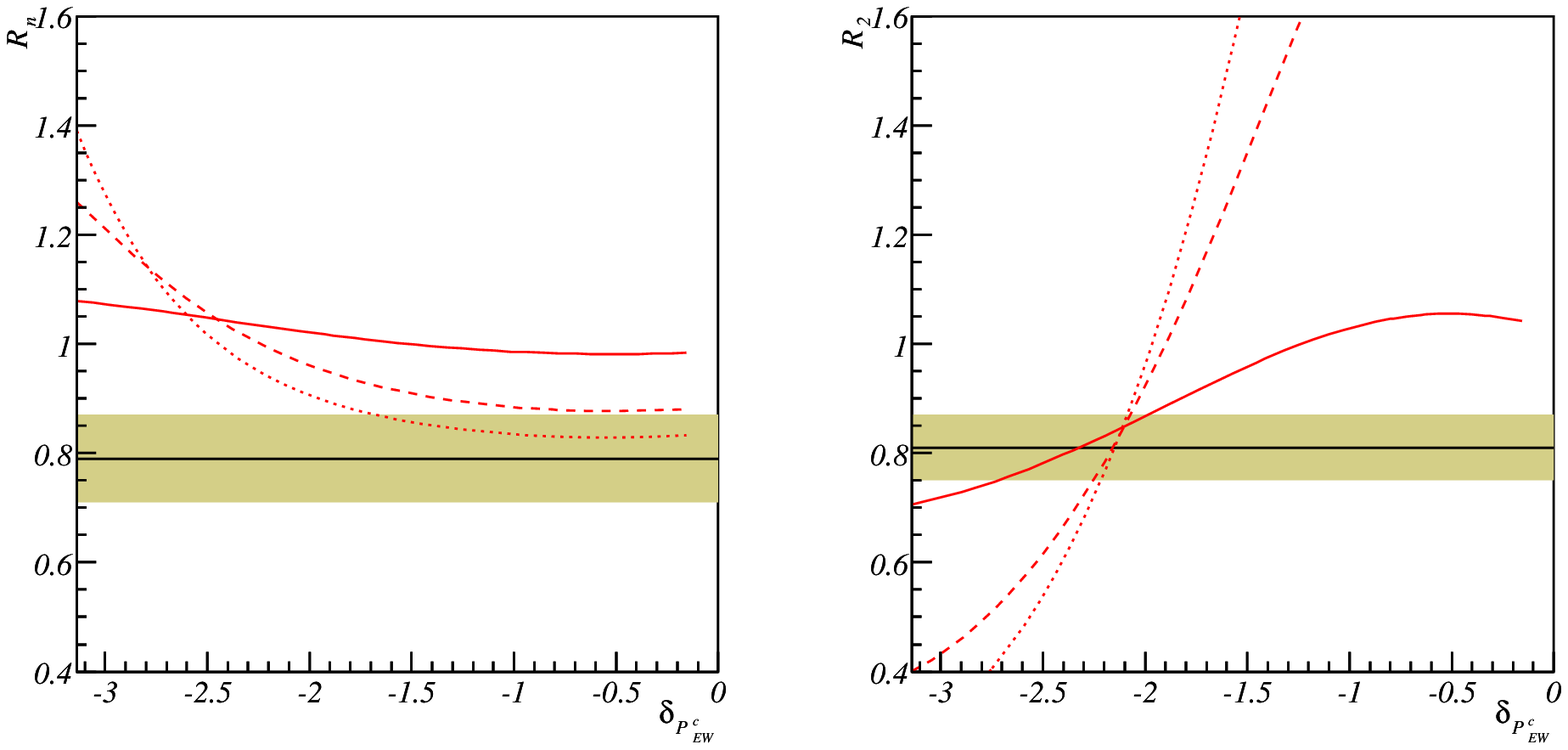}
\caption{Ratios
of $R$ and $R_2$ as functions of $\delta_{P_{\tmop{EW}}^C}$ and
$|P_{\tmop{EW}}^C|$. The three curves correspond
to  $|P_{\tmop{EW}}^C|$=0.01 (solide), 0.04 (dashed) and 0.06 (dotted)
respectively. Other parameters are fixed at
their best fitted  value in SM (according to column A of Tab.\ref{SMFit}).
The shadowed bands are the allowed range by the current
data.}
\label{pewc}
\end{center}
\end{figure}

To see the effects of $P_{EW}^C$, we give in Fig.\ref{pewc}, the
ratios of $R_n$ and $R_2$ as function of $P_{\tmop{EW}}^C$ and its
strong phase $\delta_{P_{\tmop{EW}}^C}$. In the numerical
calculations we take the values of other amplitudes from the SM
fit,  according to the first column of Tab.\ref{SMFit} or
Eq.(\ref{SA}). The values of $P_{\tmop{EW}}$ and its strong phase
are automatically generated  from Eq.(\ref{pewSM}).  It follows
from the figure that for a small $\abs{P^C_{EW}}\simeq 0.1$ the
predicted value of $R_n$ is far above the data for all values of
$\delta_{P^C_{EW}}$. To account for the current data
$\abs{P^C_{EW}}$ has to be greater than $\sim 0.4$. The strong
phase $\delta_{P^C_{EW}}$ receives strong constraint from $R_2$.
For $\abs{P^C_{EW}}$ in the range $0.1\sim 0.6$, a large negative
value of $\delta_{P^C_{EW}}=-0.2\sim -2.5$ is favored.

Including $P^C_{EW}$ as a new free parameter while keep Eq.(\ref{pewSM}),
we find
\begin{align}\label{eq:pewcFit}
  \abs{P_{\tmop{EW}}^C} &= 0.025\pm 0.021  ,
 &\delta_{P_{\tmop{EW}}^C} &= -2.24^{+ 0.21}_{- 0.63}
\intertext{and  }
  \abs{P_{\tmop{EW}}} &= 0.03^{+0.02}_{-0.013}  ,
 &\delta_{P_{\tmop{EW}}} &= 0.52^{+ 0.31}_{- 0.72} ,
\end{align}
with a $\chi^2/d.o.f = 11.3/10$. The best fits of other parameters are listed
in the first column of Tab.\ref{sublead}.  Clearly, there is a strong
cancellation in the sum of $P_{\tmop{EW}} + P_{\tmop{EW}}^C$ as
required by Eq.(\ref{pewSM}).  It is of interest to note that for both
tree and penguin diagrams the color suppressed diagrams are not
necessarily suppressed. Furthermore, the current data suggest that
\begin{align}
\abs{\frac{C}{T}}\simeq \abs{\frac{P^C_{EW}}{P_{EW}}}\approx 0.8
\end{align}
Thus the relative enhancements are likely to be  universal. This is  again in favor
of the conjecture that it has a strong interaction origin which is flavor independent.
Comparing with Eq.(\ref{eq:pewcFit}),
one sees that $P^C_{EW}$ is on the lower side to account for the suppression of $R_n$.
The best fitted ratios are $R_n\simeq 0.89$ and $R_2 \simeq 0.78$ respectively. Thus a large
$P^C_{EW}$ improves the agreement with the data.
%
%In this case  $P_{\tmop{EW}}$ is found to be $0.028^{+
%0.018}_{- 0.012}$ and $\delta_{P_{\tmop{EW}}} = 0.51^{+ 0.33}_{- 0.75}$.

%
% fit with both pew and pewc free
%
%
Taking both $P_{EW}$ and $P_{EW}^C$ as independent free parameters, we
get the following fit result
 \begin{align}\label{eq:pewcFit}
  \abs{P_{\tmop{EW}}^C} &= 0.016\pm 0.02  ,
 &\delta_{P_{\tmop{EW}}^C} &= -2.59^{+ 0.4}_{- 1.7}
\intertext{and  }
  \abs{P_{\tmop{EW}}} &= 0.027^{+0.016}_{-0.014}  ,
 &\delta_{P_{\tmop{EW}}} &= 0.69^{+ 0.3}_{- 0.6} ,
\end{align}
with $\chi^2_{min}/d.o.f=6.6/8$. The color suppressed electroweak
penguin is found to be reduced but still significant. In this
case, all the $\pi K$ ratios are well reproduced. Note that the
best fits correspond to $|\hat{P}_{EW}/\hat{T}|=0.032\pm 0.018$
which again implies a violation of the SM relation.

Finally, we emphasize that  the large $P_{\tmop{EW}}^C$ within
the SM may distinguish itself from the one beyond the SM by
the prediction of direct CP violation in $B \rightarrow \pi^- \pi^0$ which
should be vanishing in the former case and small but nonzero in the later.

%%%%%%%%%%%%%%%%%%%%%
%\subsection{New physics effects}
%%%%%%%%%%%%%%%%%%%%%

%comments on new physics
Given a small $P_{EW}^C$ relative to $P_{EW}$, the current data
may  imply new physics beyond the SM.  New physics models
significantly contributing to charmless $B$ decays may include
various SUSY models \cite{He:2004wh}, $Z'$ models from extra
$U(1)$ gauge symmetry \cite{He:1999az,Barger:2004hn} and
two-Higgs-doublet models (2HDMs)\cite{Xiao:2002mr}. Among various
versions of two-Higgs-doublet models, the general 2HDM with
spontaneous CP violation can provide rich sources of CP violation
\cite{wu:1994ja,wolfenstein:1994jw,wu:1999fe,zhou:2000ym,Wu:2001vq}
and significant corrections to the electro-weak penguin through
charge or neutral Higgs boson exchanges.  The models with 4th
generation may also give sizable contributions. For model with
both two-Higgs-doublet and 4th generation quark, the effects could
be more significant through neutral-Higgs
loops\cite{Wu:1998ng,Wu:2004kr,Wu:2005qx}.

%s4
%%%%%%%%%%%%%%%%%%%%
%
\section{nonfactorizable diagrams}\label{nonfac}
%
%%%%%%%%%%%%%%%%%%%%
We now go a step further to discuss the effects of other subleading
nonfactorizable diagram such as $E$, $A$ and $P_A$. They are
expected to be very small from factorization based estimations. However,
in the presence of large FSI, there could be mixing among diagrams which
may enhance the sizes of subleading diagrams
\cite{Wolfenstein:1995zy,Neubert:1997wb,Gerard:1997kv}.
In
view of the current puzzling pattern of the data, the possibility of
anomalously large nonfactorizable diagrams can not be excluded \cite{Feldmann:2004mg}.

Due to the limited number of data points, it is not possible to directly extract
all of them simultaneously from the current data of $\pi\pi$, $\pi K$ and $K K$.
Instead, to obtain the typical
sizes of those diagrams, we shall consider several typical cases, in each case
only one of the diagrams is assumed to be dominant  while the other two are
small. For simplicity, we only consider the case where $P_{EW}$ is fixed as
the SM value.

%%%%%%%%%%%%%%%%%%%
\subsection{$W$-exchange diagram $E$}
\begin{figure}[htb]
\begin{center}
\includegraphics[width=0.95\textwidth]{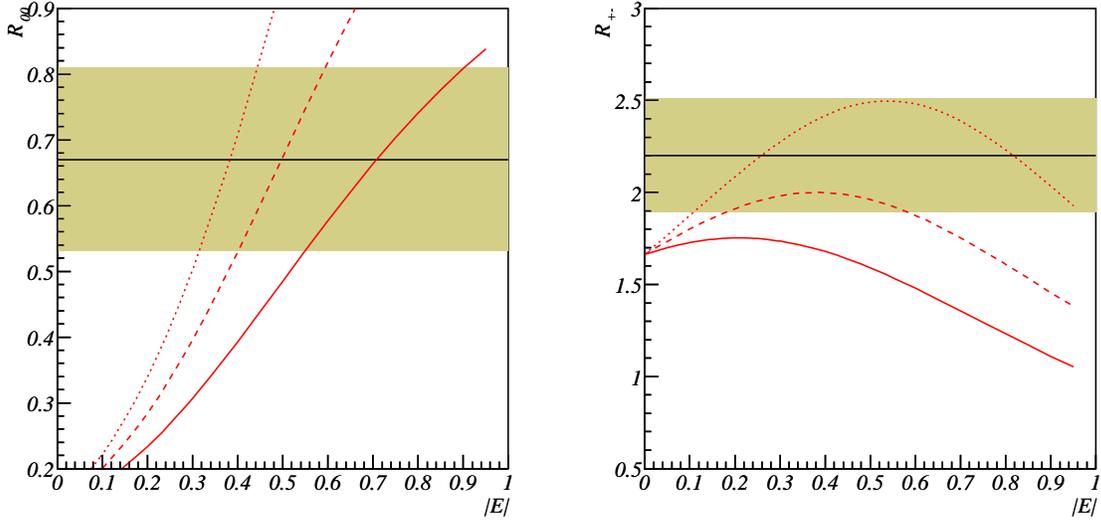}
\caption{$R_{00}$ and $R_{+ -}$ as functions of $E$ with different values of $\delta_E$
The three curves correspond to $\delta_E =$1.8 (solid), 2.0 (dashed), 2.2(
dotted). The shadowed bands are the experimentally $1\sigma$ allowed ranges. }
\label{e}
\end{center}
\end{figure}

%%%%%%%%%%%%%%%%%%%
It has been argued that sizable $E$ with constructive (destructive) interference with
$C(T)$ can help to understand the large value of $\abs{C / T}$ obtained when
$E$ is absent\cite{Buras:2003dj,Buras:2004ub,Buras:2004th}.
Since the main contribution to
$\pi^+ \pi^-$ is from $T + E$ while it is $C - E$ for $\pi^0 \pi^0$
mode. To illustrate  how $E$ improves the consistency with the $\pi\pi$
data with a small $\abs{C / T}$, we fix the tree and color-suppressed tree
to be
\begin{align}
|T| = 0.9, \quad |C| = 0.3, \quad \text{and} \quad \delta_C = 0 ,
\end{align}
respectively. The QCD
penguin is fixed at $|P| = 0.1$, $\delta_P = - 0.5$ and $P_{\tmop{EW}}$
is fixed at its SM value. In this case, the dependence of $R_{00}$ and $R_{+ -}$
with $E$ and its strong phase $\delta_E$ is plotted in Fig.\ref{e}

As shown in the figure, for a small ratio of $\abs{C / T}$=0.3,  the data require a
large $|E| \simeq 0.3 \sim 0.5$ with a large  strong phase of $\sim 2.0$. 
However, The
origin of large $E$ is  still a challenge for theory.
%Note that, although a large
%$E$ provides a solution to the puzzle in $\pi \pi$ modes, it does not help to
%reduce  the  large $\abs{C' / T'}$ independently obtained from $\pi K$ modes.

Assuming $E$ is dominant, we find a big value from a fit to the data
\begin{eqnarray}\label{bestFitE}
  \abs{E} = 0.46^{+ 0.26}_{- 0.31} & , & \delta_E = 2.86^{+ 0.17}_{- 0.23}.
\end{eqnarray}
The whole fit result is given in the second column (B) of Tab.\ref{sublead}. The
$\chi^2_{min}/d.o.f$ is 13.2/10.  Note that although $\abs{C / T}$ is reduced to
$\sim 0.28$, the relative strong phase $\delta_C$ is found to be large
$\delta_C=-2.1^{+0.83}_{-0.72}$ in contradiction with factorization based estimates.
Thus only introducing a large $W$-exchange diagram will  not be enough to
coincide with factorization results.

% comments on Ptu
An alternative way to reduce $C/T$ is to make $P_{tu}$ significantly different
than $P$ or $P_{tc}$. This is less likely  as $P_u$ is greatly suppressed
relative to $P_t$ by small $u$-quark mass in Wilson coefficient. The
important FSI process such as charming penguin only affects $P_{tc}$. A fit taking $P_{tu}$ and $P_{tc}$ as independent
parameters shows that the best fitted $P_{tc}$ is close to the factorization estimate
of $P$ while $P_{tu}$ is large and compatible with $T$ in size \cite{Chiang:2004nm},
which is quite unreasonable.

% comments on pi K modes
As it was mentioned previously, an enhanced $E$ has no effect in
$\pi K$ modes, thus can not solve the $\pi K$ puzzle and explain
the obtained large $C'/T'$. The best fitted branching ratios in
$\pi K$ still exhibit the puzzling patterns.

%%%%%%%%%%%%%%%%%%%%%
\subsection{Annihilation diagram $A$}
%%%%%%%%%%%%%%%%%%%%%

\begin{figure}[htb]
\begin{center}
\includegraphics[width=0.95\textwidth]{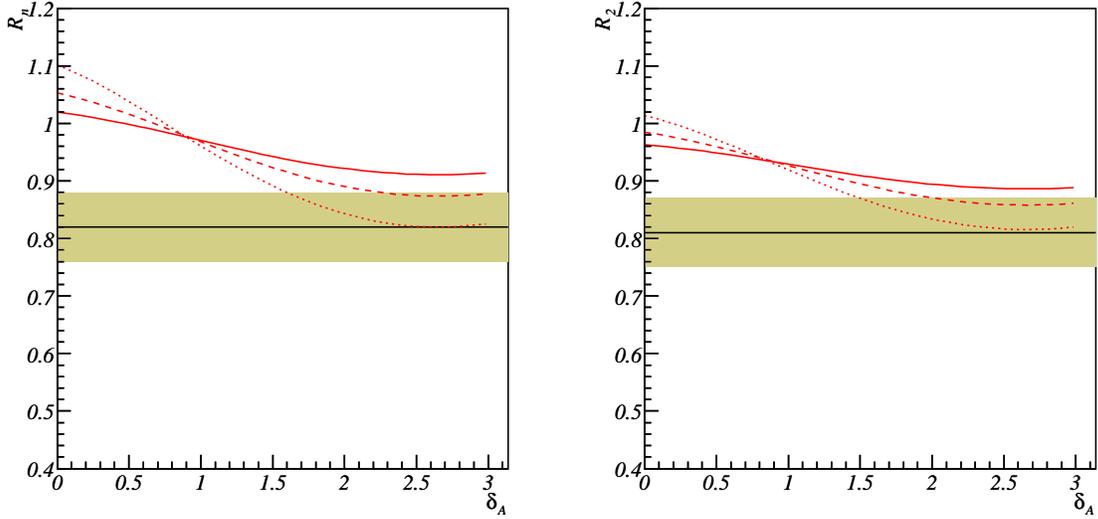}
\caption{$R$ and $R_2$ as function of $\delta_A$ with different values of $A$. The
  three curves correspond to $A =$0.3 (solid), 0.5 (dashed), 0.8(dotted).The
  shadowed bands are the experimentally allowed ranges. }
\label{a}
\end{center}
\end{figure}

The annihilation diagram $A$ appears in $\pi^- \overline{K^{}}^0$ and
$\pi^0 K^-$ modes.  The current data indicate that both of them are large in
comparison with  $\pi^+ K^-$ mode, which is characterized by the suppression
of $R_2$ and $R$.
This may imply a sizable $A$ in these modes as discussed in the previous sections.
%The effects of $A$ is manifest
%in the following figures (Fig.\ref{a}) in which
In Fig.\ref{a} the two ratios are given as functions of $A$ and
$\delta_A$. The other parameters are fixed at the SM best fit
value in Tab.(\ref{SMFit}).

One sees from the figure that
the two ratios have a similar behavior under the changing of $A$
and $\delta_A$. For $A$ ranges between $0.3 \sim 0.8$, both ratios
decrease with $\delta_A$ growing. For $\abs{A}=0.5\sim 0.8$ the
curves  fall down to the experimentally allowed ranges at
$\delta_A > 2.0$. Since both ratios contain the same  term
$\mathcal{A}+\mathcal{P}$, they have similar dependences on $A$
and $\delta_A$. This will lead to a cancellation for the ratio of
the ratios: $R / R_2$=$2 \tmop{Br} ( \pi^0 K^- ) / \tmop{Br} (
\pi^- \bar{K}^0 )$ should be  very close to unity, which agrees
with the data well.

Assuming $A$ is dominant  over other subleading diagrams 
and using the scenario A for SU(3) breaking, 
we find
\begin{eqnarray}
  |A| = 0.23^{+0.12}_{-0.09} & , & \delta_A = 2.77^{+0.28}_{-0.35} ,
\end{eqnarray}
with $\chi^2_{min}/d.o.f=13.8/10$. The whole fitting results are
listed in column (D) of Tab.\ref{sublead}.  One sees that the best
fitted value of $A$ is moderate, which helps to reduce $R$ and
$R_2$ but is not large enough to reproduce the central values of
the two ratios.  Note that even this value of $|A|\sim 0.2$ is
still much larger than that from factorization estimation which is
suppressed by a factor of $f_B/m_B$.

\begin{table}[htb]
\begin{center}\begin{ruledtabular}
\begin{tabular}{lllll}
                    &  A                         & B                       & C                       & D \\ \hline
$|T|$               &$0.51\pm0.027$              &$0.94^{+0.31}_{-0.24}$   &$0.53\pm0.026$           &$0.51\pm0.027$           \\
$|C|$               &$0.47\pm0.048$              &$0.26^{+0.2}_{-0.11}$    &$0.48^{+0.1}_{-0.046}$   &$0.48\pm0.044$           \\
$\delta_C$          &$-0.97^{+0.28}_{-0.22}$     &$-2.1^{+0.83}_{-0.72}$   &$-1.1^{+0.2}_{-0.55}$    &$-1.1^{+0.2}_{-0.18}$    \\
$|P|$               &$0.094^{+0.0018}_{-0.0021}$ &$0.097\pm0.0022$         &$0.094\pm0.0015$         &$0.093\pm0.0015$         \\
$\delta_P$          &$-0.41\pm0.14$              &$-0.26^{+0.087}_{-0.13}$ &$-0.47^{+0.093}_{-0.11}$ &$-0.54^{+0.1}_{-0.13}$   \\
$\abs{P_{EW}}$      &$0.03^{+0.019}_{-0.013}$    &$0.011\pm0.0011$         &$0.011\pm0.0011$         &$0.011\pm0.0011$         \\
$\delta_{P_{EW}}$   &$0.52^{+0.31}_{-0.71}$      &$-0.28^{+0.14}_{-0.16}$  &$-0.54^{+0.11}_{-0.25}$  &$-0.53\pm0.11$           \\
$\abs{P_{EW}^C}$    &$0.025^{+0.018}_{-0.02}$    & 0(fix)                  & 0(fix)                  &0(fix) \\
$\delta_{P_{EW}^C}$ &$-2.24^{+0.21}_{-0.61}$     & 0(fix)                  & 0(fix)                  &0(fix) \\
$|E|$               &         0(fix)             &$0.46^{+0.26}_{-0.31}$  & 0(fix)                  &0(fix)\\
$\delta_E$          &         0(fix)             &$2.86^{+0.17}_{-0.23}$  & 0(fix)                  &0(fix)\\
$\abs{P_A}$         &         0(fix)             &           0(fix)        &$0.035^{+0.026}_{-0.15}$&0(fix)\\
$\delta_{P_A}$      &         0(fix)             &           0(fix)        &$-2.26\pm0.48$            &0(fix)\\
$\abs{A}$           &         0(fix)             &           0(fix)        &0(fix)                   &$0.23^{+0.12}_{-0.087}$\\
$\delta_{A}$        &         0(fix)             &           0(fix)        &0(fix)                   &$2.77^{+0.28}_{-0.35}$ \\
$\gamma$            &$0.93^{+0.13}_{-0.16}$      &$0.94^{+0.12}_{-0.15}$   &$0.93^{+0.13}_{-0.43}$   &$0.93^{+0.12}_{-0.15}$   \\
$\chi^2_{min}/d.o.f$&11.3/10                     &13.2/10                  &14.3/10                     &13.8/10                     \\
\end{tabular}\end{ruledtabular}
\end{center}
\caption{Global fits to $\pi\pi$, $\pi K$ and $KK$ data with subleading diagrams. The four columns corresponds
to the four cases in each of them one subleading diagram is set to be free parameters}
\label{sublead}
\end{table}

%%%%%%%%%%%%%%%%%%%%%%%%%%%
\subsection{Penguin annihilation diagram $P_A$}
%%%%%%%%%%%%%%%%%%%%%%%%%%%

In $\pi \pi$ modes, the penguin induced annihilation diagram $P_A$
contributes to only low isospin final states ($I=0$) and  it
always comes together with QCD penguin $P$ in $\pi \pi$ modes.  Although
$P_A$ is often neglected in the literature, its effects are
however effectively absorbed into QCD penguin in $\pi \pi$ modes.
Thus the QCD penguin extracted from $\pi \pi$ modes are effectively
$\tilde{P} = P + P_A$.  In general $P_A$ acquires a strong
phase different than that of $P$, namely the strong phase of
$\tilde{P}$ and $P$ are different.  Since there is no SU(3) counter
part $P'_A$ appearing in $\pi K$ modes, $P'$ is still have the same
strong phase as that of $P$ in $SU(3)$ symmetry. This
introduces an effective SU(3) breaking in strong phase
between $\tilde{P}$ extracted from $\pi\pi$ modes and $P'$ from $\pi
K$. A fit to the current data gives the following values
\begin{eqnarray}
  P_A = 0.035^{+ 0.026}_{- 0.015} & , & \delta_{P_A} = -2.26\pm 0.48 .
\end{eqnarray}
Thus its size is compatible with that of electro-weak penguin. The
$\chi^2_{min}/d.o.f$ is found to be 14.3/10. The best fitted other
parameters can be found in the column (C) of Tab.\ref{sublead}.
The best fitted value corresponds to $\tilde{P} \simeq 0.093$ and
$\delta_{\tilde{P}} \simeq -0.85$, which corresponds to a strong
phase shift of
\begin{align}
\Delta\delta_P=\delta_P'-\delta_{\tilde{P}}\simeq 0.38.
\end{align}
This is in a good agreement with the previous analyses in
Refs.\cite{Wu:2002nz,Wu:2004xx}.

%s5

%%%%%%%%%%%%%%%%%%%%%
%
\section{Implications for $K K$ modes}\label{KKmodes}
%
%%%%%%%%%%%%%%%%%%%%%

The $K K$ modes are much more sensitive to the subleading diagrams
$\mathcal{E}$, $\mathcal{A}$ and $\mathcal{P}_A$.
The $K^0\bar{K}^0$ mode is dominated by QCD penguin as
discussed in section \ref{noSU3}.
The decay amplitudes of the other two modes read
\begin{eqnarray}
  \bar{\mathcal{A}} ( K^+ K^- )
& = & - \left[ \lambda_u ( E'' - P''_A ) -
  \lambda_c P''_A ] \right. ,
\nonumber\\
  \bar{\mathcal{A}} ( K^- \bar{K}^0 )
& = & \lambda_u ( - P'' + \frac{1}{3}
  P^{C''}_{\tmop{EW}} + A'' ) - \lambda_c ( P'' - \frac{1}{3} P^{C''}_{\tmop{EW}} ) .
\end{eqnarray}
The $K^+ K^-$ modes depend only on the subleading diagrams $E$ and
$P_A$.  Thus it provides an ideal avenue to explore their effects.
From the current upper bound of $\tmop{Br} ( K^+ K^- ) \lesssim 1.8$\cite{hfag},
and  the SU(3) relation of scenario A,
the exchanging diagram $E$ receives a constraint of
\begin{align}
|E|\alt 0.3,
\end{align}
which limits its contribution to $\pi\pi$ modes to be moderate as
the best fit to $\pi\pi$ modes require an $|E|\approx 0.48$ in
Eq.(\ref{bestFitE}). It is expected that stronger constraint on
$E$ will be found with more precise data in the near future.

%A
%large $E \approx 0.3$ will leads to large $K^+ K^-$ closing to the
%current experimental bonds of $\tmop{Br} ( K^+ K^- ) \lesssim 1.8$.
The direct CP violations for $K^+ K^-$ and $K^-\bar{K}^0$ reads
\begin{align}
  &a_{\tmop{CP}} ( K^+ K^- )  \nonumber\\
 &=  \frac{2| \lambda_u \lambda_c | |P''_A E''| \sin \gamma
  \sin \delta}{| \lambda_u |^2 ( |E''|^2 + |P''_A|^2 - 2 |P''_A E''| \cos \delta )
+ | \lambda_c |^2 |P''_A|^2 + 2| \lambda_u \lambda_c | |P''_A | \cos \gamma
( |E''|\cos \delta - |P''_A| )}
\intertext{and}
&a_{\tmop{CP}} ( K^- \bar{K}^0 )  \nonumber \\
&\simeq \frac{2 |\lambda_u\lambda_c| |P'' A''| \sin\gamma\sin\delta'}
{|\lambda_u|^2 (|A''|^2+|P''|^2-2 |A'' P''| \cos\delta')
+|\lambda_c|^2 |P''|^2 +2 |\lambda_u\lambda_c| |P''|
\cos\gamma(|A''| \cos\delta'-|P''|)
}
\end{align}
where $\delta = \delta_{P''_A} - \delta_{E''}$ and  $\delta'=\delta_{P''}-\delta_{A''}$.
In the expression of $a_{\tmop{CP}} ( K^- \bar{K}^0 )$ the color suppressed
electroweak penguin are neglected. A nonzero
$a_{\tmop{CP}} ( K^+ K^- )$ will definitely indicate both nonzero $E$ and
$P_A$.
%The magnitude  of  $a_{\tmop{CP}} ( K^+ K^- )$ depends on both $r$ and
%$\delta$.
In spite of the small branching ratio, in the case that $E$ and
$P_A$ are compatible in size, and the strong phase difference is
large, then the direct CP violation could be significant.

%Taking the typical value of $P = 0.09$ obtained
%from the prevous sections, one finds
%\begin{eqnarray*}
%  \tmop{Br} ( K^0 \bar{K}^0 ) & \approx & \tmop{Br} ( K^- \bar{K}^0 ) \approx
%  0.6
%\end{eqnarray*}
%which is slight lower but still in agreement with the current $K^0 \bar{K}^0$
%data.
Unlike in the $\pi K$ modes where $A$ is  suppressed by a factor $|\lambda_u^s
/ \lambda_c^s |=\mathcal{O}(\lambda^2)$, in $K^- \bar{K}^0$ mode, it is not
suppressed. Thus it is promising to probe $A$ in  $K^-
\bar{K}^0$ mode. A sizable annihilation diagram  $A$ will show up either
through  the  difference between $\tmop{Br} ( K^0 \bar{K}^0 )$ and $\tmop{Br}
( K^- \bar{K}^0 )$ or through the nonzero direct CP violation i.e.
$a_{\tmop{CP}} ( K^- \bar{K}^0 ) \neq 0$.

\begin{figure}[htb]
\begin{center}
\includegraphics[width=0.95\textwidth]{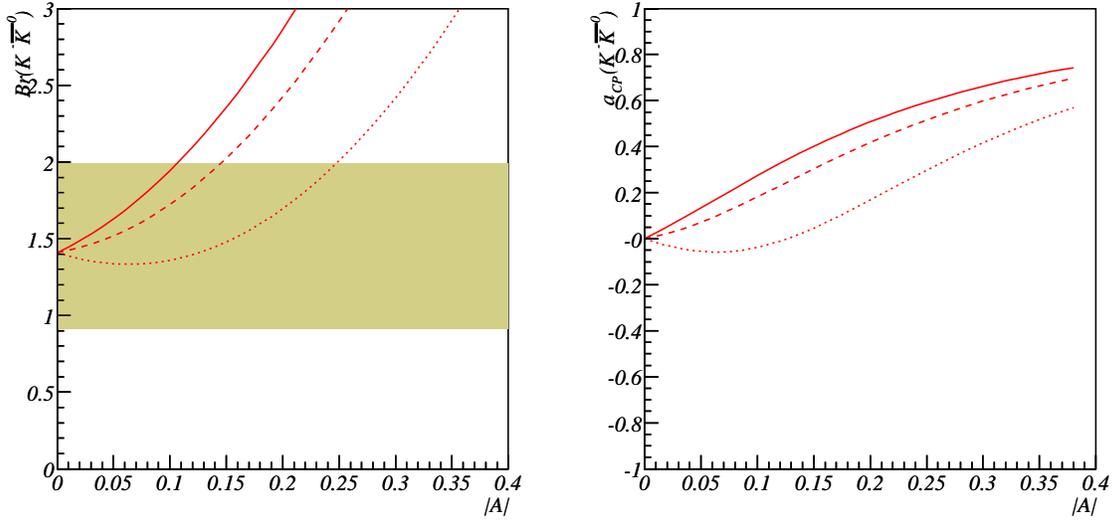}
\caption{
  Branching ratio and direct CP violation of $K^-\bar{K}^0$ as functions of $A$
  with different values of $\delta_{A}$. The three curves correspond to $\delta_{A}
  =$0 (solid), 2.5 (dashed), 3.0(dotted) respectively.  The shadowed band is the
  experimentally $1\sigma$ allowed range.  Other parameters are fixed at their best fitted
  value in SM (according to column A of Tab.\ref{SMFit}). The SU(3) breaking effects 
  are taken into account according to scenario A.
%$R$ and $R_3$ as function of $\delta_A$ with different values of $A$. The
}
\label{aKK}
\end{center}
\end{figure}
In Fig.\ref{aKK}, the decay rate and direct CP violation of
$K^-\bar{K}^0$ mode are plotted as function of $A$. In the
numerical calculations, the value of $A"$ and $P''$ is calculated from the
best fitted value of $P$ according to scenario A. For
$\delta_{A}$ in the range $0\sim 2.5$, both decay rate and
direct CP violation increase with $A$ increasing. One sees from
the figure that for $|A|=0.15\sim 0.2$ and $\delta_{A}=2.5\sim
3.0$, $a_{CP}(K^-\bar{K}^0)$ can reach $0.2\sim 0.4$ with the
branching ratio in agreement with the current data. Thus with
significantly large subleading diagrams, it is promising to
observe large direct CP violation in this decay mode.
%
%
%For $A\sim 0.2$,  $a_{CP}(K^-\bar{K}^0)$
%can reach $0.2 \sim 0.4$, for $\delta_A=0\sim 2.5$.  Therefore it is promising to discover
%large CP violation in this mode with the future high precision data.

%%%%%%%%%%%%%%%%%
%
\section{Summary}\label{summary}
%
%%%%%%%%%%%%%%%%%
In summary, we have systematically studied charmless $B$ decays $B\to \pi\pi, \pi K$
and $KK$, following a strategy that making independent analysis for $\pi\pi$, $\pi K$, $K K$
modes individually  as the first step and then connecting them through various
SU(3) relations.  The separated analysis allowed us to clarify  the origins
of the inconsistence or puzzles  revealed by the current data.
Independent analysis on $\pi \pi$ and $\pi K$ modes $both$ favor
large ratio of $C/T$ and $C'/T'$ with large strong phases, which
suggests that they are more likely to originate from long distance
strong interactions rather than large non-factorizable exchange
diagrams $E$. The sizes of QCD penguin diagrams in $\pi\pi$, $\pi
K$ and $KK$ are independently extracted and were found to follow a
pattern of SU(3) breaking in a good agreement with factorization
estimation $P'/P\simeq P''/P'\simeq f_K/f_\pi$.
Global fits to these modes have been carried out under various
scenarios of SU(3) relations. All the results show good
determinations of weak phase $\gamma$ in consistency with the
Standard Model (SM) and prefer a large electroweak penguin $(
P_{\tmop{EW}} )$ relative to $T + C$ with a large strong phase.
Within the SM, it may require an enhancement of color suppressed
electro-weak penguin ($P_{\tmop{EW}}^C$) with destructive
interference to $P_{\tmop{EW}}$. The possibility of the presence
of new physics effects can not be excluded.
We have also investigated the possibility of sizable contributions
from nonfactorizable diagrams such as $E$, $A$ and $P_A$. Their
sizes could be significantly larger than the expected ones. The
typical sizes of $|E|$ could reach to $|E| \approx 0.3$ as required
by the $\pi\pi$ and $K K$ data, $|A|$ could reach to $|A| \approx 0.2$ while
$|P_A|$ has a typical value of $ |P_A| \approx 0.03$.
The sizable subleading diagrams may change significantly the
predictions for the yet to be seen $K^+K^-$ and $K^-\bar{K}^0$
modes. The CP violation in $K^-\bar{K}^0$ modes could reach $a_{CP}
\simeq 0.2\sim 0.4$ for a large value of $A$ and $\delta_A$ in the
range of $|A|=0.15\sim 0.2$ and $\delta_{A}=2.5\sim 2.5$.
It would be encouraging to expand the investigation of subleading diagrams to
decay modes involving $\eta$ and $\eta'$ final states. Although these decay
modes receive significant contributions from additional flavor singlet penguin
diagrams \cite{Gronau:1996ng,Dighe:1996gq,Dighe:1997hm,Chiang:2001ir} or
nonstandard contributions through
$\bar{c}c$\cite{Halperin:1997as,Halperin:1998ma} or QCD anomaly
\cite{Fritzsch:1997ps,Fritzsch:2003pt}, as more data points are involved,
stronger  constraints on the subleading diagrams will be expected.  Thus it
will enable us to test the SM using the full diagrammatic decomposition.

%%%%%%%%%%%%%%%%%%%%%%%%%%%-Reference-%%%%%%%%%%%%%%%%%%%%%%%%%%%%%%%%%%%%%%%%%%
\begin{acknowledgments}
YLW  is supported in part by the key project of NSFC and
Chinese Academy of Sciences. YFZ is grateful to S. Safir for
reading the manuscript and useful comments.
\end{acknowledgments}

\bibliographystyle{apsrev}
\bibliography{/home/zhou/reflist/reflist,misc}
%\bibliography{reflist,misc}

%%%%%%%%%%%%%%%%%%%%%%%%%%%%%%%%%%

\end{document}